\documentclass[
  journal=pasa,
  manuscript=research-paper
]{cup-journal}

\usepackage{microtype,siunitx,booktabs,amssymb,amsmath,xcolor,multirow,longtable,pdflscape,threeparttablex,bm,dblfloatfix}
\usepackage{caption}
\usepackage{subcaption}
\definecolor{link}{rgb}{0,0,1}
\usepackage[pdfnewwindow=true,
            colorlinks=true,
            linkcolor=link,
            citecolor=link,
            filecolor=link,
            urlcolor=link, 
            ]{hyperref}
\usepackage{xurl}  
\newcommand{\zHD}[1]{\ensuremath{z_{\text{HD}}^{\mathrm{#1}}}}

\newcommand\zhel{\ensuremath{z_{\text{hel}}}}

\newcommand{\ten}[1]{\ensuremath{10^{#1}}}
\newcommand{\tten}[1]{\ensuremath{\times 10^{#1}}}
\newcommand{\kmsMpc}{km~s$^{-1}$~Mpc$^{-1}$}
\newcommand{\hMpc}{$h^{-1}$Mpc}
\newcommand{\kms}{km~s$^{-1}$}

\title{WiFeS observations of nearby southern Type Ia supernova host galaxies}

\author{Anthony Carr}
\affiliation{School of Mathematics and Physics, University of Queensland, Brisbane, QLD 4072, Australia}
\alsoaffiliation{Korea Astronomy and Space Science Institute, Yuseong-gu, Daedeok-daero 776, Daejeon 34055, Republic of Korea}
\email[Anthony Carr]{anthonycarr@kasi.re.kr}

\author{Tamara M.\ Davis}
\affiliation{School of Mathematics and Physics, University of Queensland, Brisbane, QLD 4072, Australia}

\author{Ryan Camilleri}
\affiliation{School of Mathematics and Physics, University of Queensland, Brisbane, QLD 4072, Australia}

\author{Chris Lidman}
\affiliation{The Research School of Astronomy and Astrophysics, Australian National
University, Stromlo, ACT 2601, Australia}

\author{Kenneth C.\ Freeman}
\affiliation{The Research School of Astronomy and Astrophysics, Australian National
University, Stromlo, ACT 2601, Australia}

\author{Dan Scolnic}
\affiliation{Department of Physics, Duke University, Durham, NC 27708, USA}

\doi{10.1017/pasa.XXXX.XX}

\received {dd Mmm YYYY}
\revised  {dd Mmm YYYY}
\accepted {dd Mmm YYYY}
\published{dd Mmm YYYY}


\begin{document}

\begin{abstract}
We present high-resolution observations of nearby ($z\lesssim0.1$) galaxies that have hosted Type Ia supernovae to measure systemic spectroscopic redshifts using the Wide Field Spectrograph (WiFeS) instrument on the Australian National University 2.3 m telescope at Siding Spring Observatory. 
While most of the galaxies targeted have previous spectroscopic redshifts, we provide demonstrably more accurate and precise redshifts with competitive uncertainties, motivated by potential systematic errors that could bias estimates of the Hubble constant ($H_0$).
The WiFeS instrument is remarkably stable; after calibration, the wavelength solution varies by $\lesssim 0.5$ \AA\ in red and blue with no evidence of a trend over the course of several years.
By virtue of the $25\times 38$ arcsec field of view, we are always able to measure the redshift of the galactic core, or the entire galaxy in the cases where its angular extent is smaller than the field of view, reducing any errors due to galaxy rotation.
We observed 185 southern SN Ia host galaxies and measured the redshift of each via at least one spatial region of a) the core, and b) the average over the full-field/entire galaxy.
Overall, we find stochastic differences between historical redshifts and our measured redshifts on the order of $\lesssim \ten{-3}$ with a mean offset of 4.3\tten{-5}, and normalised median absolute deviation of 1.2\tten{-4}.
We show that a systematic redshift offset at this level is not enough to bias cosmology, as $H_0$ shifts by $+0.1$ \kmsMpc\ when we replace Pantheon+ redshifts with our own, but the occasional large differences are interesting to note.
\end{abstract}

\section{Introduction}
\label{sec:WiFeSintro}
Given that the discrepancy between the Planck 2018 CMB measurement of $H_0$, $67.4\pm0.5$ \kmsMpc{} \citep{Planck2018} and most recent SH0ES measurement, $73.04\pm1.04$ \kmsMpc{} \citep{Riess2022} is now at the 5$\sigma$ level, we must ensure we have a comprehensive understanding of any possible systematic errors.
In the case of local distance ladder supernova cosmology, these systematics can take many forms when measuring Cepheid/SN brightnesses, recession velocities and distances, and may bias our measurements of cosmic expansion.
An extensive set of systematics was recently explored as part of the recent SH0ES/Pantheon+ collaboration \citep[][and references therein]{Brout2022cosmo,Riess2022}, including, but not limited to the geometric--Cepheid distance calibration sample \citep{Yuan2022}; the Cepheid--SN calibration sample and Cepheid metallicity dependence \citep{Riess2022}; SN photometry calibration \citep{Brout2022cals}; SN dust and colour \citep{Popovic2021a}; SN peculiar velocities \citep{Peterson2022}; and SN redshifts \citep{Carr2022}.
The general conclusion from these analyses is that SN systematics are not a solution to the Hubble tension, as each individual systematic can only realistically account for a small fraction of the tension, and in fact, often increases the tension.

The most straightforward of the above systematics to test are the redshifts (and, for similar reasons, peculiar velocities; however, since these are modelled or measured using the redshifts, we do not study them here).
A systematic shift to redshift can easily influence $H_0$ in much the same way as a systematic shift to measured SN magnitudes, as a magnitude shift is degenerate with a shift in $H_0$.
A shift in redshift of, e.g.\ 1\tten{-4}, would be equivalent to a magnitude shift of around 9 mmag at $z=0.0233$ and 1.5 mmag at $z=0.15$. 
The effect is smaller at higher redshift due to the sub-linear nature of the distance modulus--redshift relation. 
For the same reason, downward shifts in redshift have a slightly larger effect on magnitude than the same shift upward.
According to \citet{Davis2019}, redshift errors on the order of only 5\tten{-4} can bias $H_0$ by nearly 1 \kmsMpc, if the errors are systematic and at low-$z$ (the smaller the redshift, the larger the effect).

In the case of real data, as part of the Pantheon+ analysis, \citet{Carr2022} studied the effects of redshift errors on SN cosmology.
The goal of \citet{Carr2022} was to overhaul the redshifts used in supernovae analyses, particularly at low-$z$ where SNe Ia are rare, and we currently rely on a vast collection of historical data.
While no observations were conducted, the redshifts were improved in multiple ways: host associations were checked, higher quality literature redshifts were sourced where possible, uncertainties were studied, the transformation from the measured redshifts to the CMB-frame was corrected, and the peculiar velocity model was improved.
Ultimately, they showed the combined effects of existing redshift and peculiar velocity errors amounted to a negligible shift in $H_0$ of $-0.12\pm0.20$ \kmsMpc.
While these redshift systematics have now been thoroughly ruled out as being a complete solution to the Hubble tension from historical data, there remains a possibility that errors in the measurements of the redshifts are still present.

\citet{Carr2022} used historical data, so here we measure new redshifts to test whether systematics in the historical data are present.
Specifically, we target bright, nearby galaxies, which are the most influential to $H_0$.
In addition, these galaxies have the most potential to be biased by, e.g.~pointing errors, where a spectroscopic slit or fibre may be placed not on the core, but elsewhere in the galaxy (such as the location of the supernova). 
Historically, SN surveys have used long-slit spectroscopy to observe the galactic core at the same time as the SN, which requires precise alignment of the slit.
Host redshifts are also sometimes measured from the SN classification spectrum if the emission lines are bright enough.
In both cases, the rotation of the galaxy will bias the redshift to some extent if the host redshift is not measured from the core, i.e.~if the slit angle is misaligned or the fibre does not cover the core.
To combat the effects of potential observational bias, we use integral-field spectroscopy to ensure we can capture the systemic redshift.

The paper is set out as follows: in Section~\ref{sec:WiFeSobs}, we detail the target selection, observation and reduction process for our program. 
In Sections~\ref{sec:WiFeScal} and \ref{sec:WiFeSwlcal} we detail our analysis of the calibrations we undertook and the performance of the instrument over the course of our program.
Finally, we describe our redshift results compared to historical data and their impact on cosmology in Section~\ref{sec:WiFeSresults}, and then conclude in Section~\ref{sec:WiFeSconclusion}.

\section{Observations and data} \label{sec:WiFeSobs}
We start with a description of the overall strategy we took for our observation program, with Section~\ref{subsec:strat} describing the technical set-up and rationale.
In Section~\ref{subsec:targets}, we detail the target selection and observation strategy.
We then move to the data reduction process in Section~\ref{subsec:reduc} and the post-processing we perform in the form of spatial binning in order to measure the high-quality redshifts in Section~\ref{sec:WiFeSzs}.

\subsection{The WiFeS instrument}\label{subsec:strat}
We take advantage of integral-field spectroscopy to measure the spatial variation in redshift across the face of large galaxies and gather enough signal-to-noise (S/N) to successfully redshift smaller/fainter galaxies.
We use the Wide Field Spectrograph (WiFeS) instrument mounted on the Australian National University 2.3 metre telescope (ANU 2.3m) at Siding Spring Observatory (SSO). 
The field-of-view is $25\times38$ arcsec, and in our operation mode, each spaxel is $1\times 1$ arcsec.
See \citet{Dopita2007} for the full instrument specifications and \citet{Dopita2010} for the measured performance.

\begin{table}[!t]
    \centering
    \scriptsize
    \caption{WiFeS grating wavelength ranges and dispersion, as measured from our reduced data. Wavelength limits are rounded to the nearest 5 \AA. The beamsplitters are named after their dichroic wavelength split in nanometres. The dispersion and upper wavelength limit for each grating varied over time by less than 0.02\%.}
    \label{tab:WiFeS_specs}
    \begin{tabular}{lcccc}
    \toprule
     & & $\lambda_{\mathrm{min}}$ & $\lambda_{\mathrm{max}}$ & Dispersion \\
     Grating & Beamsplitter & \AA & \AA & \AA/pixel \\
    \midrule
    U7000 & RT480 & 3290 & 4355 & 0.272 \\
    B7000 & RT615 & 4180 & 5550 & 0.347 \\
    R7000 & RT480 & 5290 & 7025 & 0.439 \\
    I7000 & RT615 & 6830 & 9055 & 0.566 \\
    B3000 & RT560 & 3200 & 5900 & 0.775 \\
    R3000 & RT560 & 5300 & 9565 & 1.25 \\
    \bottomrule
    \end{tabular}
\end{table}

\begin{figure}[!t]
    \centering
    \includegraphics[width=0.99\textwidth]{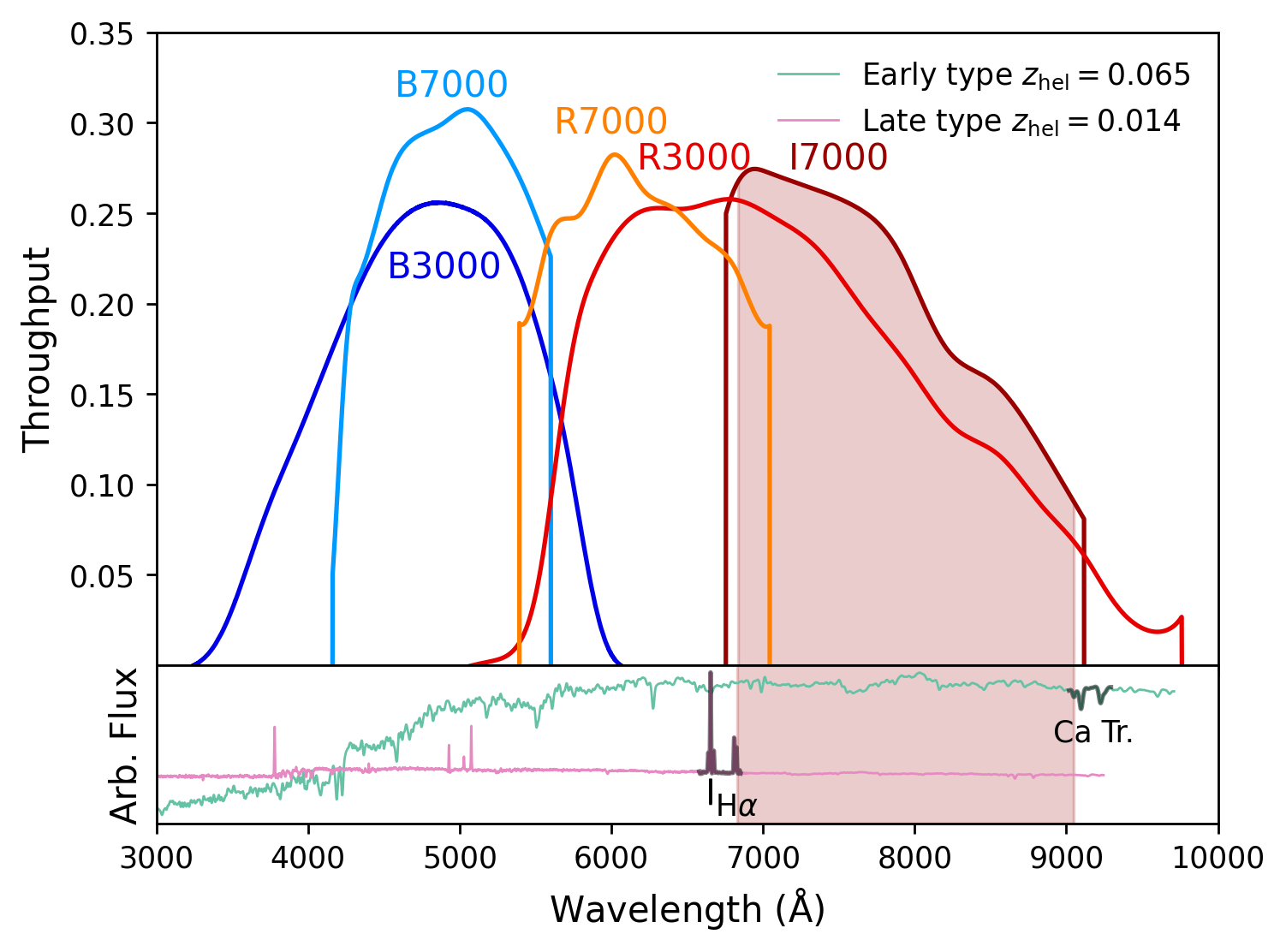}
    \caption{Measured throughput of each grating except for U7000 from \citet{Dopita2010}. The shaded region represents the I7000 wavelength bounds in practice (Table~\ref{tab:WiFeS_specs}). The bottom panel shows two examples of spectra that one might fail to measure a redshift from the I7000 grating. For a redshift of $\lesssim0.014$ ($\gtrsim0.065$), none of the H$\alpha$ (Calcium triplet) region is present in I7000 leaving only weak or no features.}
    \label{fig:throughput}
\end{figure}

We originally planned to observe using only $R=7000$ gratings, since these offer higher precision redshifts at the cost of reduced wavelength coverage compared to the $R=3000$ gratings.
The specifications of the grating suite are described in Table~\ref{tab:WiFeS_specs}, and the throughput curves measured by \citet{Dopita2010} are shown in Figure~\ref{fig:throughput}.
The ANU 2.3m offers full optical spectral coverage at $R=7000$ over four individual gratings: the ultraviolet and blue gratings are paired with the red and infrared gratings respectively.
We opted to use only the B7000+I7000 grating pair as the U7000+R7000 did not offer useful spectral coverage considering observations would take twice as long and add extra overheads swapping grating pairs and beamsplitter.

However, after the first two observing runs, we switched to observing with full spectral coverage at $R=3000$, which still offers excellent precision and better S/N for the same exposure time. 
We made this decision due to the fact that at the low redshifts we were targeting, the I7000 grating was sometimes a trade-off between the Calcium triplet and H$\alpha$ region being redshifted out of spectral coverage.
We show this in Figure~\ref{fig:throughput} with two example spectra at redshifts where the I7000 spectral region would be devoid of features.
If we have good enough S/N for a late-type galaxy, or the galaxy has a particularly strong Calcium triplet, we would still be able to identify features in I7000 below $z\approx 0.014$, and similar for early-type galaxies above $z\approx 0.065$.
However, for the final two observation runs, we observed using the B3000+R3000 gratings since they offer full spectral coverage with a generous overlap, and as such have no such restrictions on where typical optical galaxy emission and absorption features land.

\subsection{Target Selection}\label{subsec:targets}
The catalogue was created using the Pantheon supernova sample \citep{Scolnic2018} since we conducted all of our observations before the release of Pantheon+.
Both the Pantheon and Pantheon+ supernova samples are a vast collection of SNe Ia light curves (1048 and 1701 respectively) and redshifts from different sources, both low- and high-$z$ to best constrain cosmology.
At the time of observation, Pantheon was the most powerful SN sample, so we aimed to observe as many of the bright, southern SN hosts as possible.
Galaxies were chosen to be easily observable from Siding Spring Observatory (latitude $149.06^{\circ}$, longitude $-31.27^{\circ}$), i.e.~airmass $\lesssim1.5$, and with $z\lesssim0.1$ as these are the redshifts most influential to $H_0$.

The strategy was to observe in Nod\&Shuffle mode, which results in simultaneous science and sky spectra.
The target is exposed on the science CCD pixels, then the telescope is nodded to empty sky and the charge already present is `shuffled' across the CCD so that the sky is exposed on a different set of pixels.
Sky subtraction is then just the simple case of subtracting the pure 2D sky spectrum from the observed object+sky 2D spectrum within the same CCD image during data reduction, leaving just the 2D object spectrum.
Each galaxy observation was made up of three Nod\&Shuffle cycles, with each sky and object exposure being the same length.
The sky field was chosen to be as empty as possible, and close by to reduce the time to nod between frames.

Once the targets were chosen, we calculated the rough exposure time using the WiFeS performance calculator.\footnote{\url{https://www.mso.anu.edu.au/rsaa/observing/wifes/performance.shtml}}
The aim was a generous S/N of at least 20 in the blue camera after the full 3$\times$Nod\&Shuffle cycle, which was easily obtained for the brightest galaxies.
The red camera naturally gathers more signal for the same observation time, so observation time was optimised for the blue camera.
Exposure time depends on the moon phase (all observations were done in grey or dark time), the seeing full width at half maximum (FWHM; typically 1.6 arcsec at SSO), as well as the airmass and surface brightness of the target.
The average total integration time was 485 s for an estimated average g-band surface brightness of 16 mag arcsec$^{-2}$.
Subexposure times were rounded to the nearest 30 seconds rather than attempting to save small amounts of time optimising to the nearest second.

\begin{figure}[bt!]
    \centering
    \includegraphics[width=0.7\textwidth]{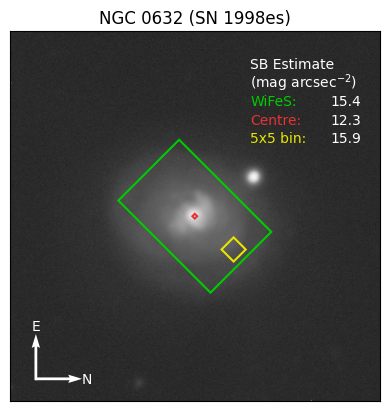}
    \caption{DECam g-band image of the galaxy NGC 632, host of SN 1998es. The three estimates of surface brightness were made over the whole WiFeS aperture, within the central pixel of the aperture, and finally, a 5x5 pixel binned aperture on the outskirts to estimate a lower limit of surface brightness.}
    \label{fig:WiFeSSBestimate}
\end{figure}

Surface brightness was estimated directly from Dark Energy Camera (DECam) images using the US National Science Foundation's NOIRLab Astro Data Lab image cutout service.\footnote{\url{https://datalab.noirlab.edu/sia.php}}
Most targets had images; for those that did not, we estimated surface brightness by comparison with similar targets that did have images.
When there were images, we opted for sky-subtracted images in the g-band as the highest priority, followed by r-band then i-band (with minor corrections to account for overestimating the g-band magnitude), and stacked images if no sky-subtracted version was available. 
The surface brightness was estimated from the images using the \texttt{photutils} python package\footnote{\url{https://photutils.readthedocs.io/en/stable/}} within different apertures, including the full WiFeS aperture, to estimate a useful average for the whole field of view.
See Figure~\ref{fig:WiFeSSBestimate} for an example.

Flux calibration stars were chosen to be CALSPEC stars\footnote{\url{https://www.stsci.edu/hst/instrumentation/reference-data-for-calibration-and-tools/astronomical-catalogs/calspec}} which are the standards used for the Hubble Space Telescope.
The only criterion for choosing a CALSPEC flux calibrator was that it was easily observable from SSO.
The CALSPEC standard stars were also used to remove Telluric absorption features from the galaxy spectra.
We trialled the use of dedicated, particularly smooth-spectrum Telluric standard stars (hot, main sequence B stars), but were unable to reliably source reference spectra.
As such, the Telluric corrections were sometimes lacking, but never resulted in a failure to compute a redshift.

Radial velocity standard stars (stars with well-known radial velocities to compare to as another form of instrument calibration) were chosen from \citet{Nidever2002}.
The stars we used were all chosen to be G- or K-type stars with a preference for giants.
They were also chosen to primarily be around $V=6$ mag, similar to the flux calibrators.

Using this strategy, we observed 213 galaxies, 185 of which were unique targets, and the rest duplicates for increased S/N. A log of our observations can be found in Table~\ref{tab:Obslog} including MJD of observation and exposure time of our main science targets and radial velocity standards.

\begin{figure*}[hp!]
  \centering
  \begin{minipage}{0.49\textwidth}  
    \centering
    \begin{subfigure}{0.49\textwidth}
      \includegraphics[clip,trim=0 0 0 22,width=\textwidth]{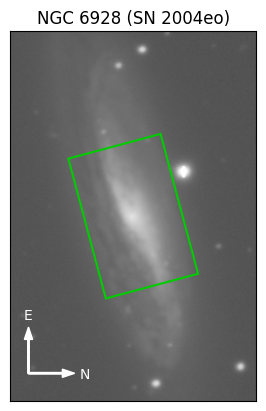}
      \caption{DECam image of NGC 6928 (r-band). }
      \label{fig:2004eosky}
    \end{subfigure}
    \hspace{6pt}
    \begin{subfigure}{0.7\textwidth}
      \includegraphics[width=\textwidth]{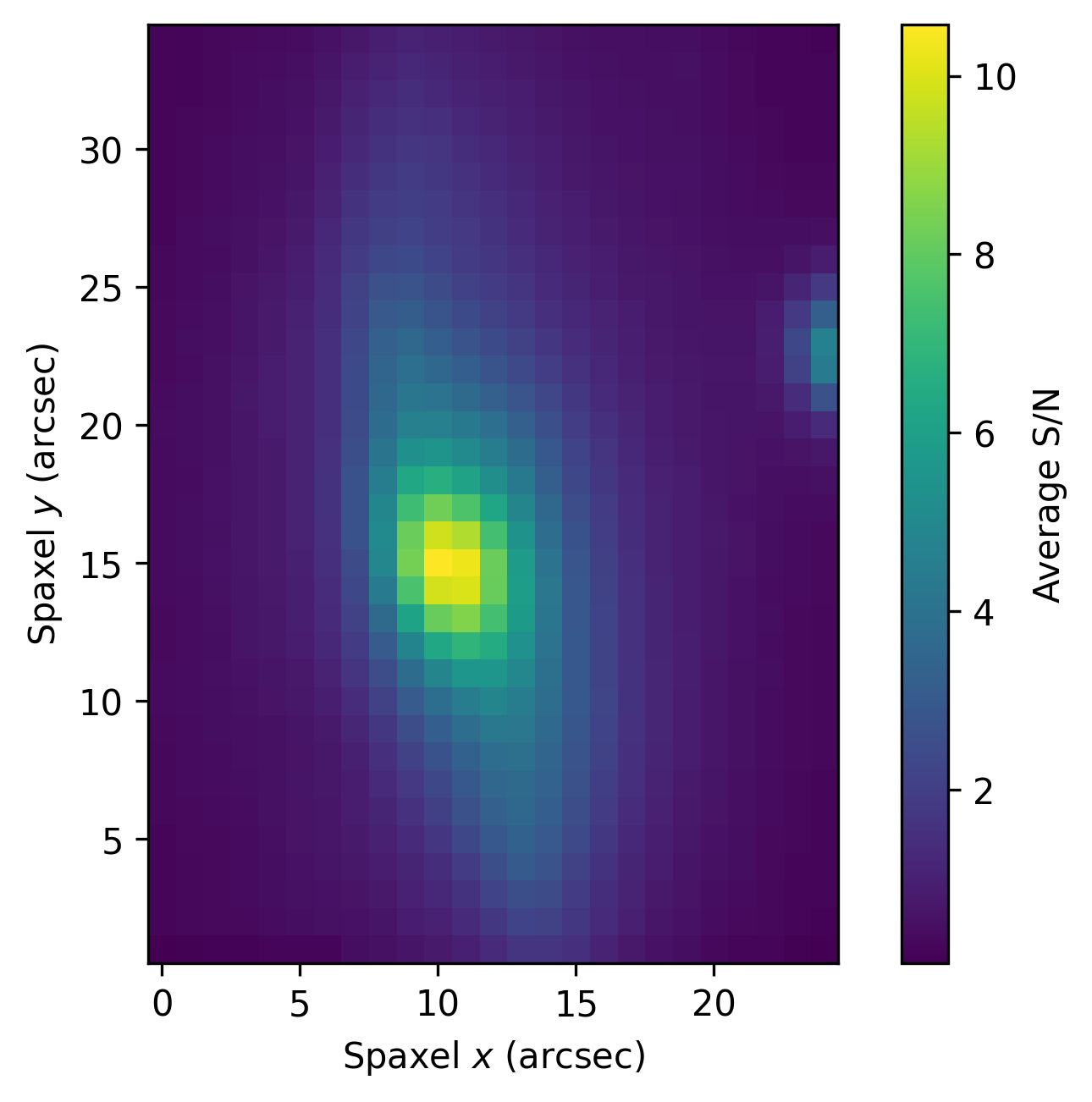}
      \caption{Average S/N per spaxel.}
      \label{fig:2004eovoronoibin}
    \end{subfigure}
  \end{minipage}
  \hfill
  \begin{minipage}{0.45\textwidth}  
    \begin{subfigure}{0.62\textwidth}
      \includegraphics[clip,trim=0 0 0 22,width=\textwidth]{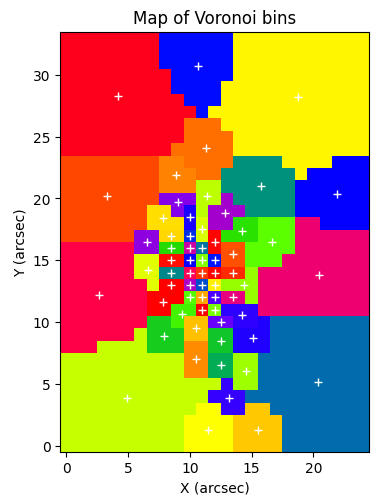}
      \caption{Optimal binning map from \texttt{vorbin}.}
      \label{fig:2004vorbinmap}
    \end{subfigure}
    \begin{subfigure}{0.8\textwidth}
      \includegraphics[width=\textwidth]{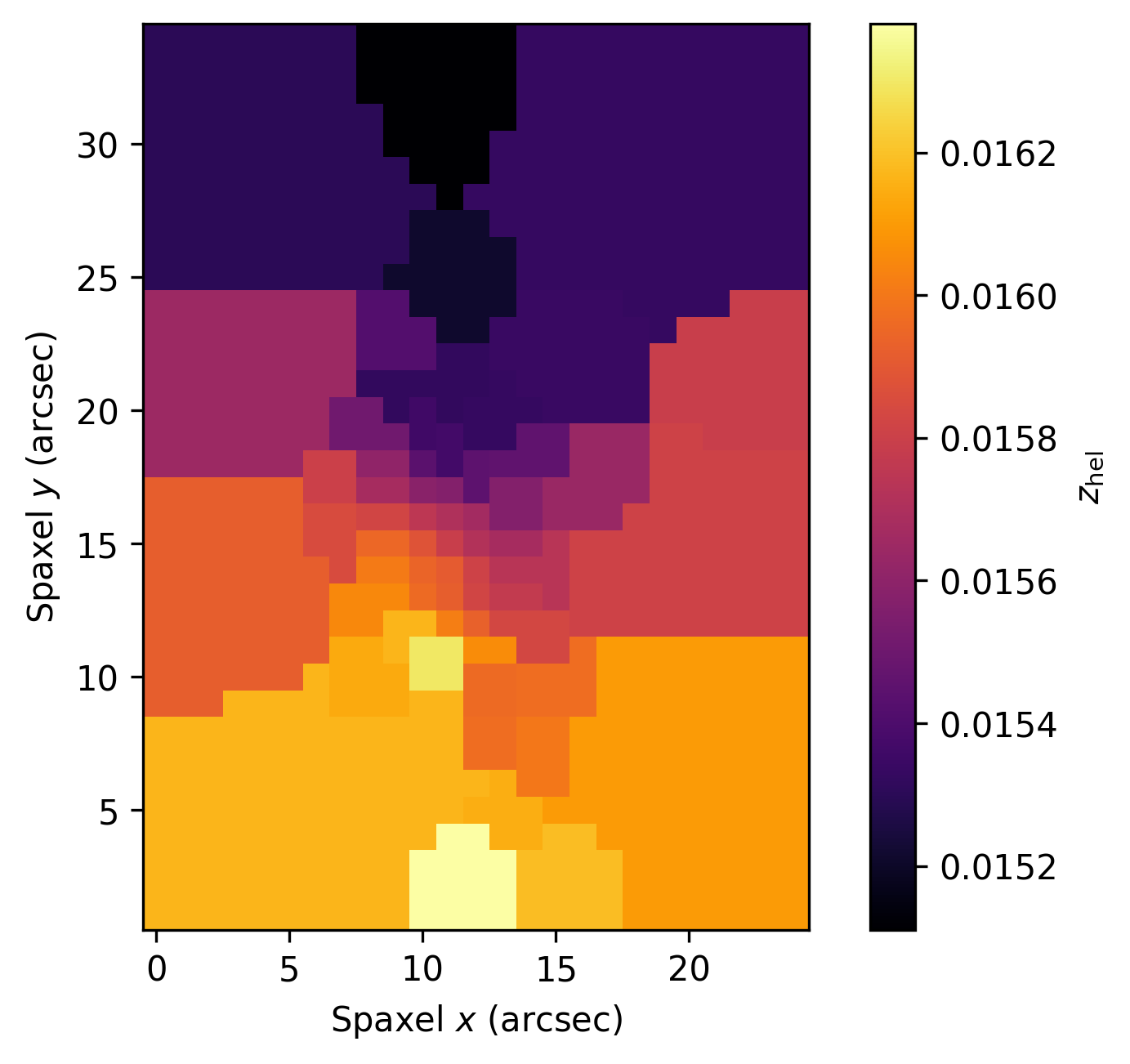}
      \caption{Redshift map.}
      \label{fig:2004eovorbinall}
    \end{subfigure}
  \end{minipage}
      \vspace{10pt}

  \begin{subfigure}[b]{0.49\textwidth}
    \hspace{10pt}
    \includegraphics[width=0.8\textwidth]{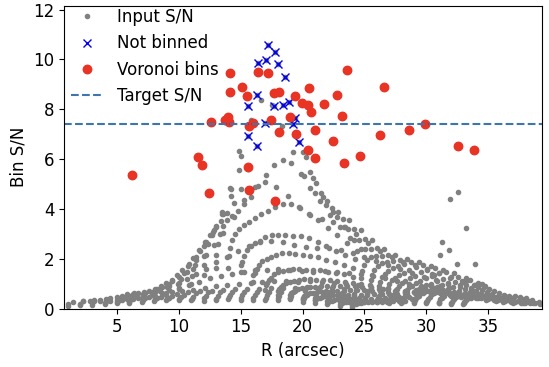}
     \caption{Binned and unbinned S/N from \texttt{vorbin}.}
    \label{fig:2004eozmap}
  \end{subfigure}
  \caption{\textbf{a)} Image of NGC 6928 on the sky with the WiFeS field of view in green. SN 2004eo occurred outside the green aperture pictured here, nearly one arcminute east-north-east of the centre of the galaxy. \textbf{b)} S/N over the WiFeS aperture, as averaged per spaxel over the entire red camera spectra. Evidence of the nearby star can be seen as an increase in S/N around spaxel coordinates (25,23), $\sim$33 arcsec from spaxel (0,0). \textbf{c)} The Voronoi binning regime that results in each bin having roughly 70\% the average S/N of the central spaxel. \textbf{d)} Final redshift map, showing rotation along the long axis. The average redshift of all the spaxel bins is $\zhel=0.01576$, with standard deviation 2.9\tten{-4}. \textbf{e)} The S/N of each unbinned and binned spaxel, with the dashed line showing the S/N target.}
  \label{fig:obs_to_z}
\end{figure*}

\subsection{Reduction}\label{subsec:reduc}
We used version 0.7.4 of the default WiFeS reduction pipeline, PyWiFeS \citep{Childress2014}, to transform the raw observations into calibrated, 3D data cubes.
In brief, the reduction pipeline pre-processes each CCD image (overscan and bias subtraction, bad pixel repair), then uses spatial calibration frames to split the data into the 25 science and 25 sky slitlets.
The instrument was designed so that the sky and science slitlets are interleaved on the detectors, and slitlets lie along detector rows.
Over the entire detector, the slitlets deviate from these rows by up to $\sim\pm0.5$ pixels.
This deviation is accounted for by observing a uniformly illuminated calibration frame obstructed by a straight wire.
The shadow of the wire is used to find the spatial zero-point of each slitlet along the $y$-direction (the long axis of the aperture) over every CCD column.

From here, the usual steps are performed: finding the wavelength solution, cosmic ray rejection, sky-subtraction, flatfielding, flux calibration and Telluric correction.
Finally, the data are reformatted to a 3D data cube.
We note that while the field of view is $25\times38$ arcsec/spaxels, in practice we trim the noise-dominated outer one or two spaxel rows (depending on the gratings and beamsplitter) so we actually use $25\times35$ spaxels for our purposes.

Due to the excellent stability of the WiFeS instrument, the wavelength solution varied only on the sub-pixel level.
We expand upon this in Section~\ref{sec:WiFeScal}.

\begin{figure*}[!t]
    \centering
        \includegraphics[width=0.32\textwidth]{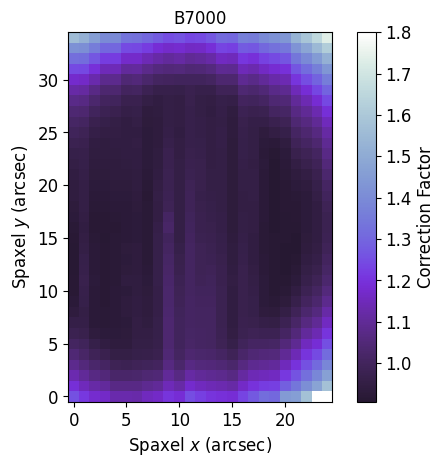}
        \includegraphics[width=0.32\textwidth]{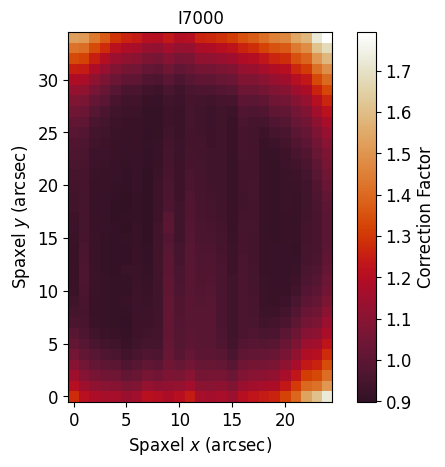}
           \hspace{20pt}
        \includegraphics[width=0.32\textwidth]{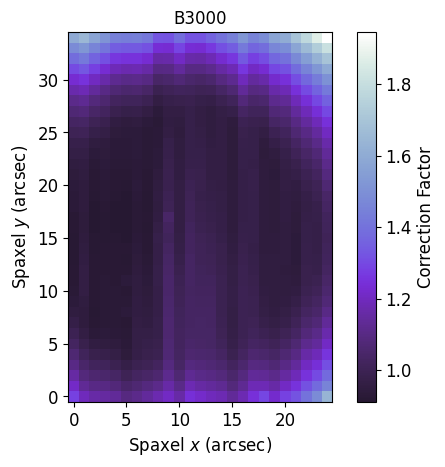}
        \includegraphics[width=0.32\textwidth]{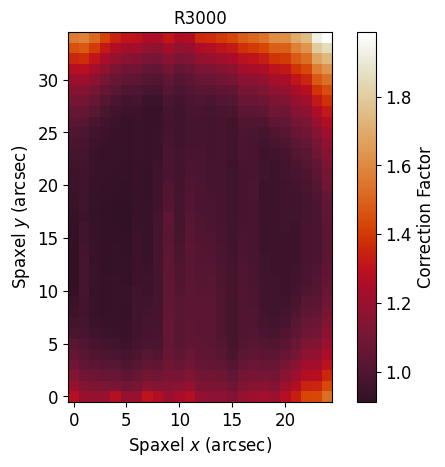}
    \caption{Illumination corrections for the red and blue camera averaged over each observation run in each operation mode, as determined from twilight flats (see text). The $R=3000$ and $R=7000$ grating corrections are basically the same since this is a spatial correction. The corrections are also comparable to those of \citet{Childress2016} indicating the stability of the instrument over several years.}
    \label{fig:WiFeSillum_corr}
\end{figure*}

\begin{figure*}[!hp]
    \centering
        \includegraphics[width=0.39\textwidth]{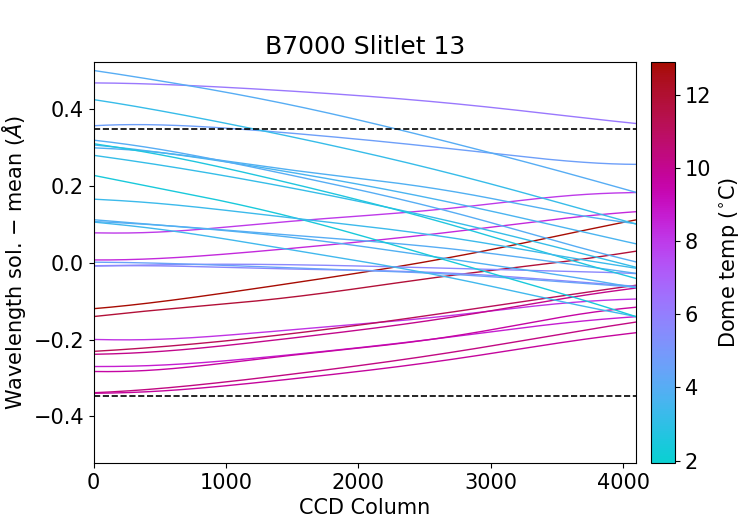}
        \includegraphics[width=0.39\textwidth]{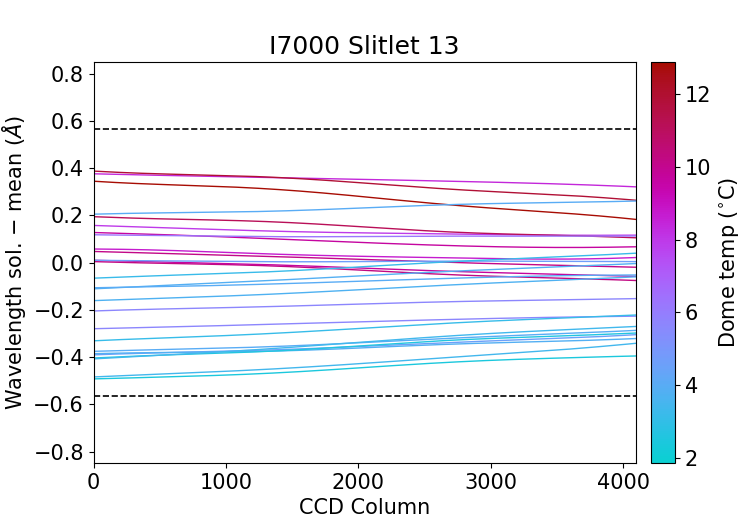}
        \includegraphics[width=0.39\textwidth]{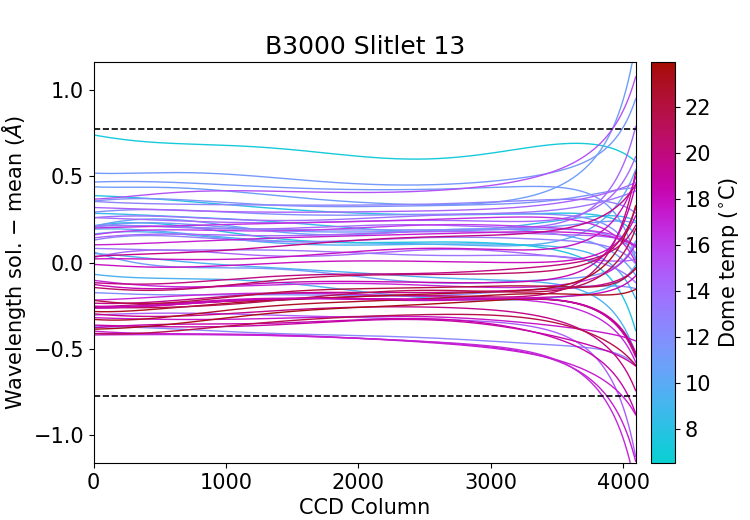}
        \includegraphics[width=0.39\textwidth]{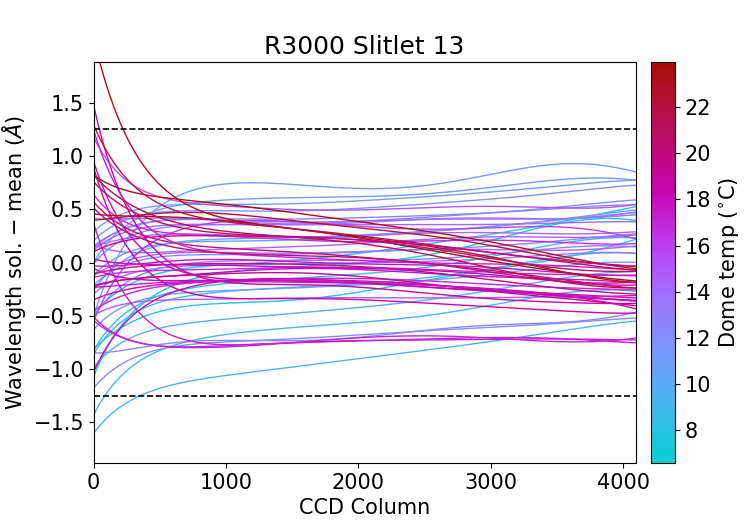}
\caption{Difference in the wavelength solution over the CCD for each resolution and camera, compared to the mean for that resolution+camera combination. We show only the middle slitlet, but the others show similar overall trends \citep[see][]{Childress2016}. Each curve is a different arc lamp observation, coloured by the temperature recorded at that time. The dashed lines denote the size of a CCD pixel. $R=7000$ shows similar levels of variation to $R=3000$ (despite the differing wavelength dispersions), except at the CCD column boundary, where the variation is less extreme because a) throughput is higher at higher resolution (especially toward the ends of wavelength coverage where B3000 and R3000 drop to near 0; see Figure~\ref{fig:throughput}), and b) the wavelength solution for $R=3000$ gratings is less stable in the dichroic region at high (low) CCD column numbers for B3000 (R300).}
    \label{fig:WiFeSwsolCCD}
\end{figure*}

\begin{figure*}[hp!]
    \centering
        \includegraphics[width=0.39\textwidth]{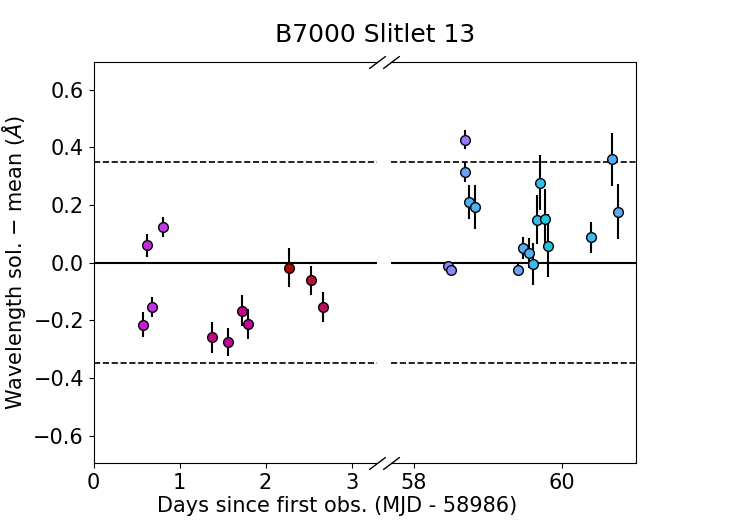}
        \includegraphics[width=0.39\textwidth]{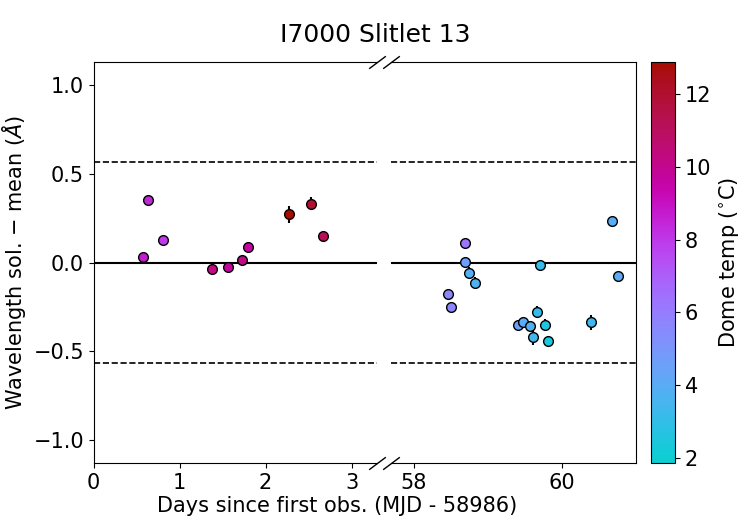}
        \includegraphics[width=0.39\textwidth]{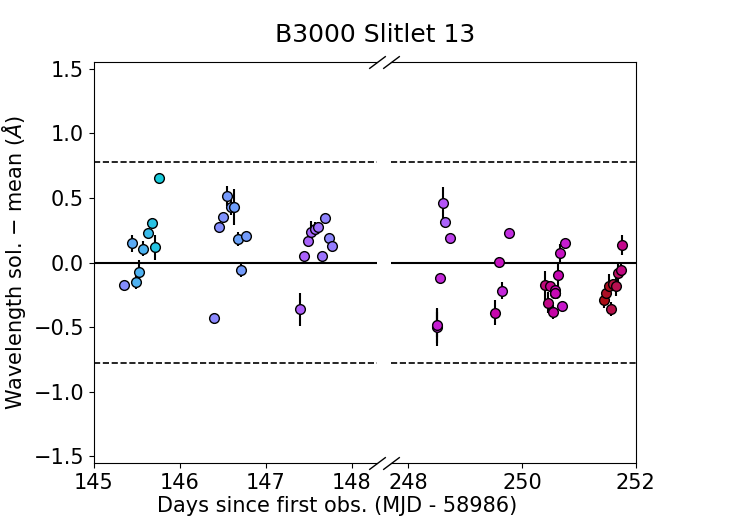}
        \includegraphics[width=0.39\textwidth]{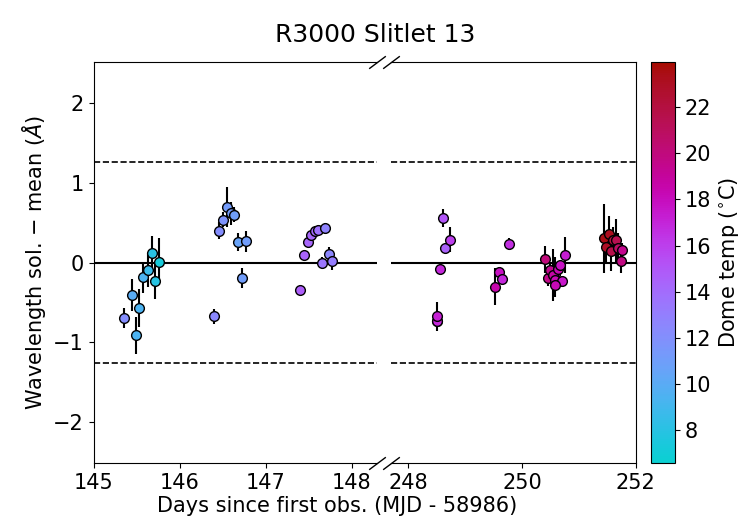}
    \caption{The same as Figure~\ref{fig:WiFeSwsolCCD} but now averaged across the CCD column and as a function of observation date. The dependence on temperature is still clear, but now the seasonal dependence is seen by the grouping in different observation runs.}
    \label{fig:WiFeSwsoldate}
\end{figure*}

\subsection{Spaxel binning and redshift measurement}\label{sec:WiFeSzs}
After reducing the data, the 3D spectral data cube contains a wealth of information.
For this work, we are mainly interested in the redshift and its spatial dependence, although there is certainly more that can be done with spatially varying, medium-resolution ($R=3000$--7000), high S/N data.
To investigate the redshift(s) of the galaxy, we first processed the data further into a format that could be ingested into the Marz redshifting tool,\footnote{\url{https://samreay.github.io/Marz}. 
The default tool does not include the calcium triplet, so we modified the source code for our own use.} which was developed primarily for the use of the Australian Dark Energy Survey and Anglo-Australian Telescope, also at SSO. 

To turn our nearly 1000 individual spaxels into a reasonable number of high S/N spectra, we spatially bin them.
This allows us to gather more signal in the outer regions of the aperture, and to successfully measure the redshift of galaxies that occupied very few spaxels.
The best tool for the purpose of binning 2D spectroscopic data is the \texttt{vorbin} python package \citep[see][]{Cappellari2003} which uses Voronoi tessellation to create bins of roughly the target S/N.
This adaptive binning method naturally creates a complete tessellation (no overlap or holes) with bins that are as compact as possible (no elongated or fragmented bins) with minimum variation in S/N.
This is due to the particular algorithm of seeding, bin-accretion and correction developed by \citet{Cappellari2003} for the purposes of integral-field spectroscopic data.

See Figure~\ref{fig:obs_to_z} for a visualisation of how observations are turned into redshifts. 
It shows the case of a galaxy larger than the aperture, where spaxels further from the bright central region are binned to a common S/N threshold. 
The redshift of each Voronoi bin is then measured separately.
From this map, the average redshift of the whole galaxy, the redshift of the centre and the redshift in the locality of the SN can all be found.
The particular example shown does not contain the SN in the aperture; there is only a moderate probability of the SN being within the field of view when centred on such large galaxies.

For most of the Voronoi binning, we set the S/N target to 90\% of the central pixel S/N by default.
This target level was sometimes adjusted in the cases where the surface brightness profile was particularly flat (requiring an increase), such as MCG-02-02-086, the host of SN 2003ic, or the galaxy as a whole was extremely bright (requiring a decrease to reduce the number of bins from $\gg$100 to $\lesssim$100), such as NGC 6928, featured in Figure~\ref{fig:obs_to_z}.
In each case, the binning was visually inspected to ensure both a decent number of bins and high-quality spectra for each.

In 24 cases, there was only one bin, necessary for the particularly small/faint/poor-quality-spectrum galaxies to successfully obtain an accurate redshift.
Where possible, we also redshift just the central region (estimated from the highest S/N spaxels), to compare with the average redshift over all Voronoi bins.
This often resulted in a strong increase in S/N; however, for the 24 cases above, a similar or lower S/N spectrum naturally resulted, since we did not use as many spaxels as for the whole galaxy.
Of the 185 galaxies we observed, 106 were at least roughly as large in the sky as the aperture, and nine were small enough to occupy only several spaxels each. 
For the standard stars, the $3\times 3$ spaxel region around the centre of the star was used for measuring redshift.

The geocentric-to-heliocentric correction to account for our motion around the Sun is automatically handled in Marz by including the relevant headers for the telescope location, observation time and object coordinates.
The distribution of geocentric corrections between $-25$ \kms\ and $+20$ \kms\ only had a slight positive gradient; however, the mean correction was 11.7 \kms\ due to a large number of corrections falling between $+20$ \kms\ and $+30$ \kms.

\section{Instrument throughput correction}\label{sec:WiFeScal}
Transforming the data from photon counts to accurate flux as a function of accurate wavelength is a multi-step process.
`Dome' flats using an internal Quartz-Iodine lamp correct for the CCD pixel-to-pixel quantum efficiency.
The Quartz-Iodine lamp is mostly `spectrally flat', in that the wavelength dependence can be removed by a moderate order polynomial, but does not illuminate the instrument perfectly uniformly.
Twilight flats using the twilit sky (which are `spatially' flat, i.e.~uniform illumination, but significant spectral deviation) are used to correct for the spatial illumination.
Once the pixel-to-pixel and large-scale illumination are corrected, one of the final steps is flux calibration, which corrects for the wavelength response and instrument+atmosphere throughput. 

\citet{Childress2016} studied the performance of the WiFeS instrument over a three-year period for the ANU WiFeS SuperNovA Programme (AWSNAP), including wavelength solution, illumination correction, and flux calibration. 
We have a similar set of data gathered over a year in two operation modes, five years later, with which to compare.
We emulate their analysis here to observe any trends over nightly to yearly scales and detect any need for manual recalibration.
We start by examining the illumination correction for our different operating modes (see Figure \ref{fig:WiFeSillum_corr}), and find no significant differences (visually) between them and \citet{Childress2016}, which speaks to the stability of the instrument.

\begin{figure*}[ht!]
    \centering
        \includegraphics[width=0.39\textwidth]{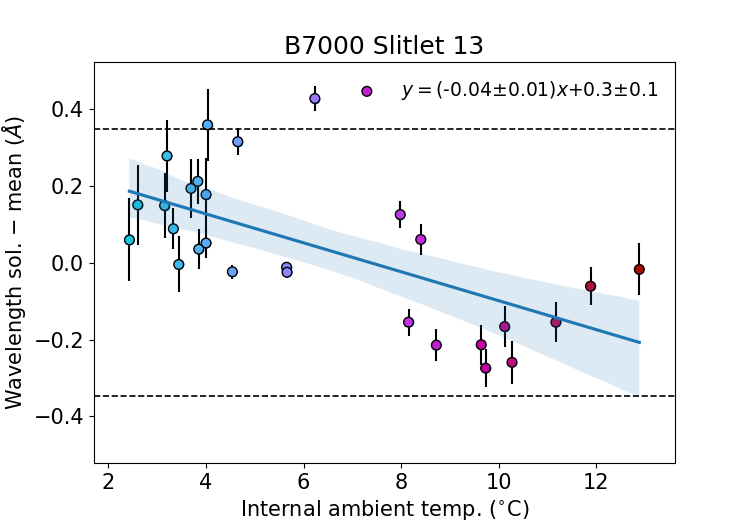}
        \includegraphics[width=0.39\textwidth]{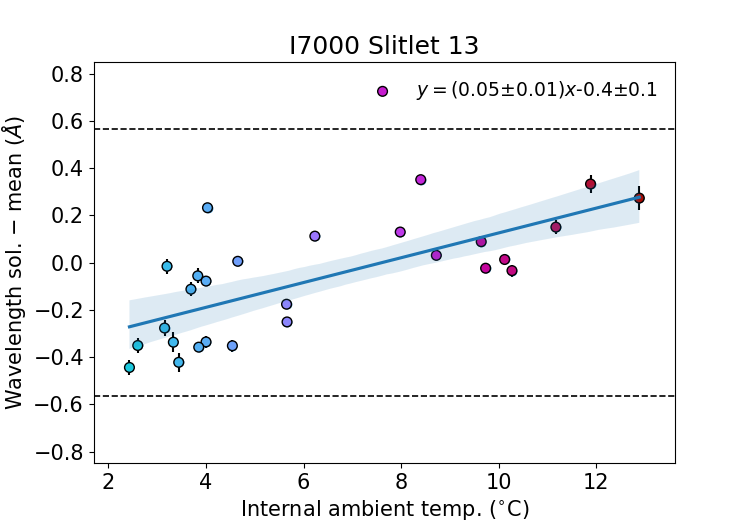}
        \includegraphics[width=0.39\textwidth]{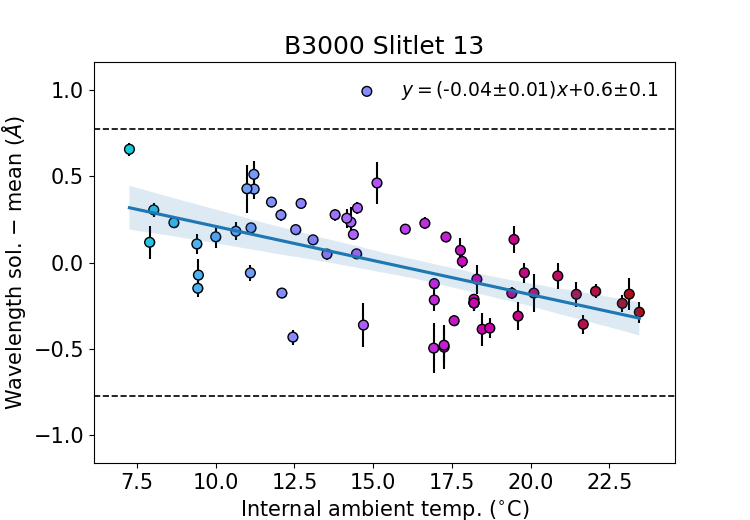}
        \includegraphics[width=0.39\textwidth]{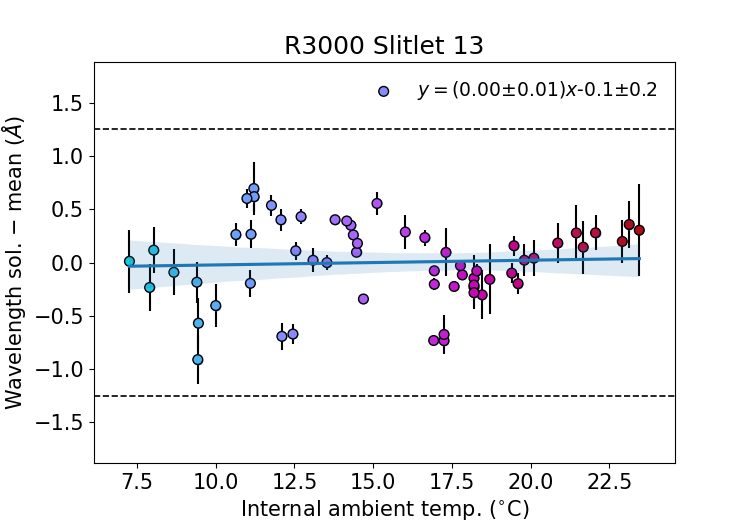}
    \caption{Average difference in wavelength solution as a function of temperature. The colour map is carried through from Figures~\ref{fig:WiFeSwsolCCD} and \ref{fig:WiFeSwsoldate}, i.e.~the $x$-axis here. Since we observed for a full year, our observations roughly capture the extremes of temperature during a typical year, so it is expected that the average wavelength solution will not vary by more than a CCD pixel. According to the linear relations fit to the temperature data, even if the temperature does vary more than expected, it would still be predominantly sub-pixel.}
    \label{fig:WiFeSwsoltemp}
\end{figure*}

\section{Wavelength calibration methods}\label{sec:WiFeSwlcal}
\begin{figure}[bt!]
    \centering
    \includegraphics[width=0.95\textwidth]{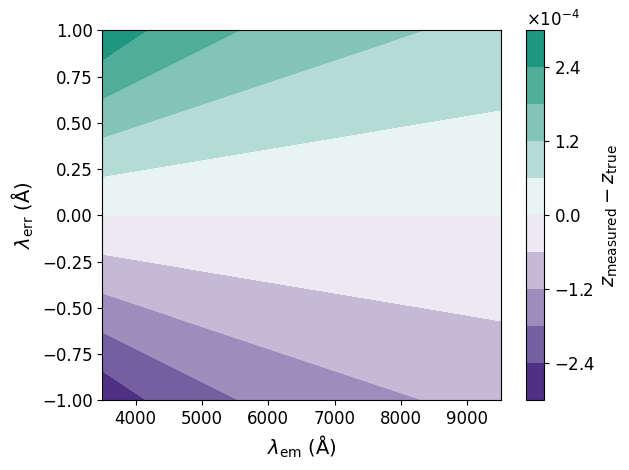}
    \caption{The redshift error resulting from measuring a spectral feature at $\lambda_{\mathrm{em}}$ up to $\pm$1 \AA\ ($\lambda_{\mathrm{err}}$) from its true value, due to, e.g.~a small residual temperature dependence in the wavelength solution. We generally redshift features in the red (H$\alpha$, Ca triplet), and the average wavelength solution error is less than 0.5 \AA\ over most of the CCD (Figure \ref{fig:WiFeSwsolCCD}), resulting in a realistic maximum redshift error of $\sim 6\tten{-5}$. Sources whose features are primarily at low wavelengths could potentially experience larger errors.}
    \label{fig:lambda_error}
\end{figure}

\begin{figure*}[bt!]
    \centering
        \includegraphics[width=0.35\textwidth]{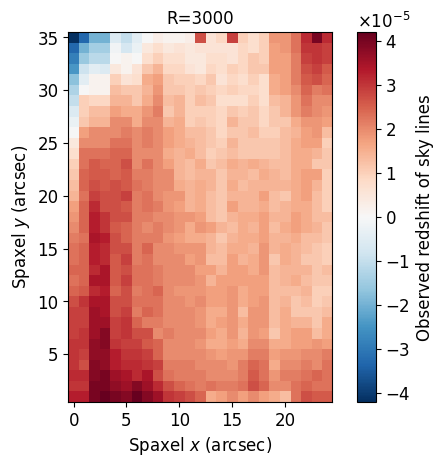}
        \includegraphics[width=0.35\textwidth]{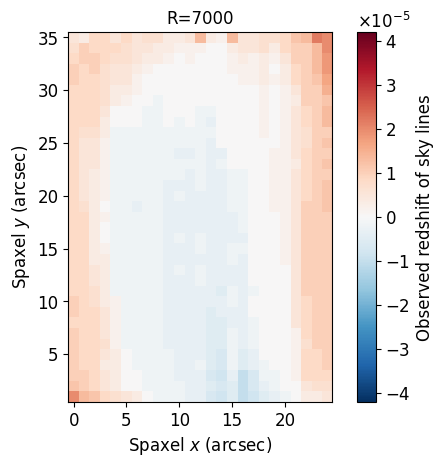}
    \caption{Wavelength solution (effectively redshift) calibration across the aperture per grating, found by means of measuring the `redshift' of the $z=0$ sky spectrum, with no spatial binning. }
    \label{fig:WiFeSskylinesols}
\end{figure*}

\begin{figure*}[bt!]
    \centering
        \includegraphics[width=0.4\textwidth]{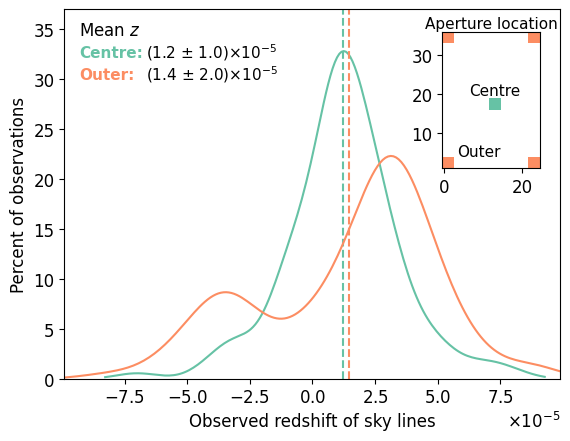}
        \includegraphics[width=0.4\textwidth]{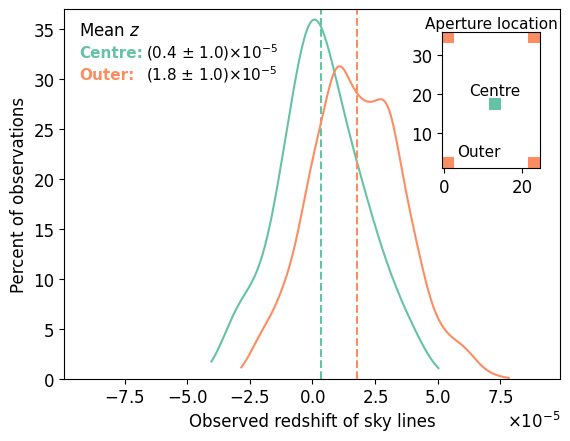}
    \caption{Wavelength solution/redshift calibration across the aperture with binning shown in the upper right of each panel, from the observed redshift of the $z=0$ sky spectrum. The left panel shows $R=3000$ and the right panel shows $R=7000$. These curves have been smoothed over the $\sim$200 sky-spectrum observations for each individual aperture region. The double-peaked nature of the $R=3000$ curve for the outer regions comes from the fact that the upper left corner is generally blueshifted, and the upper right and lower left redshifted from the centre (see Figure~\ref{fig:WiFeSskylinesols}). This is also the case for $R=7000$ but to a much lesser extent.}
    \label{fig:WiFeSskylinelocs}
\end{figure*}

In this section, we study the accuracy of the instrument so we can be assured the redshifts we measure are not biased.
We study the effects of temperature fluctuation on the wavelength solution throughout each night and over the entire observation program in Section~\ref{subsec:temp}. 
We also use the skylines that we measure simultaneously with our science targets to track how the wavelength solution varies across the aperture in Section~\ref{subsec:skylines}.
Finally, in Section~\ref{subsec:radvel}, we compare our redshift measurements of radial velocity standard stars to their accurately known values, along with the effects of the spectral template we use to measure redshift. 
In essence, we find that the wavelength solution shows excellent stability and thus our redshifts require no spatial or temperature correction.

\subsection{Arc lamp wavelength solutions and temperature dependence}\label{subsec:temp}
It is well known that temperature fluctuations inside the dome affect the wavelength solution of WiFeS spectra, as the gratings themselves thermally expand. 
We endeavoured to mitigate any temperature effects by taking frequent arc lamp calibration frames.
The response of the gratings to temperature fluctuations may be linear, which can be interpolated over easily, but the temperature fluctuations themselves are not and lag behind the dome internal temperature readings.
By making frequent arc lamp observations, temperature variations are accounted for since each science observation is calibrated using the nearest arc lamp in time, or if the science observation falls between two arc lamps, it is calibrated by the average of those wavelength solutions.
Similar to \citet{Childress2016}, we investigated the variation of the wavelength solution over the CCD, and over time, and we compare both of these to dome temperature readings to correlate with the fluctuations. 

Figures~\ref{fig:WiFeSwsolCCD}--\ref{fig:WiFeSwsoltemp} show our investigation into wavelength solution variation as measured from arc lamp observations.
Over our entire observation program that spanned one year, we used two different resolutions each for two runs (approximately spanning 6 months each).
We find very similar results to \citet{Childress2016}, in that the average wavelength solution generally deviates by $\pm 0.5$ \AA\, which for our gratings is always less than a CCD pixel (see Table~\ref{tab:WiFeS_specs}), except for a couple of B7000 measurements.
There are no long-term trends (albeit with only a single year of data) apart from the seasonal (temperature) difference.
Even extreme temperature fluctuations are expected to cause an order $\sim1$ \AA{} fluctuation in wavelength solution. 

To put into perspective how a 1 \AA\ error would appear when measuring a redshift, we show in Figure \ref{fig:lambda_error} the severity of measuring a spectral feature up to $\pm$1 \AA\ off its true value for a source $z=0.1$,\footnote{While we set the source redshift to 0.1, there is no dependency on the redshift for this additive offset to the wavelength solution. 
We could equally consider a \textit{stretch} to the wavelength solution, and this would act as an additional redshift.} over the full spectral range of WiFeS.
The error is more severe at the blue end, but we generally use features in the red for measuring redshift.
Over the entire CCD, the average temperature-induced shift is less than 0.5 \AA\, which we account for with frequent calibration, so the expected temperature-induced redshift error is much less than the maximum $\sim 0.5$ \AA\ or redshift of $\sim 6\tten{-5}$ from Figure~\ref{fig:lambda_error}.
Below, we show that when we redshift well-known objects/features, we indeed see a smaller average error.

\subsection{Skylines}\label{subsec:skylines}
Skylines are strong emission lines, mostly in infrared, that come from our own atmosphere and need to be removed from our spectra.
However, since they originate on Earth, we can use them as a test of wavelength solution in addition to arc lamps.
For every science target, we measured the `redshift' of the sky spectrum (which is coincident with but separate from the science spectra due to the Nod\&Shuffle mode of operation) in the centre and corners of the aperture to test how the wavelength solution varied across the field of view.
For a small subset of the entire science sample (one observation from every night), we did the same without any binning, i.e.~we tested the wavelength solution of every spaxel since the S/N of the individual skylines is always very high.
To get the redshift of a sky spectrum, we modified Marz to use a high S/N sky spectrum from the European Southern Observatory's \texttt{skycorr} tool \citep{Noll2014} as a template.
Figure~\ref{fig:WiFeSskylinesols} shows the average wavelength solution across the aperture, grouped by resolution.
The $R=3000$ gratings show more variation, but each only varies by $<5\tten{-5}$, and the central region is always accurate to within $\pm2\tten{-5}$.

Figure~\ref{fig:WiFeSskylinelocs} shows the results of binning the central and outer regions separately for the high and low-resolution configurations.
The $R=7000$ gratings show little variability, especially in the centre.
In every case, the mean offset is $<2\tten{-5}$, which corresponds to $<0.1$ \AA\ in our observable skyline spectral region (much less than any grating's resolution).

\subsection{Radial velocity standards}\label{subsec:radvel}
\begin{figure}[bt!]
    \centering
        \includegraphics[width=0.79\textwidth]{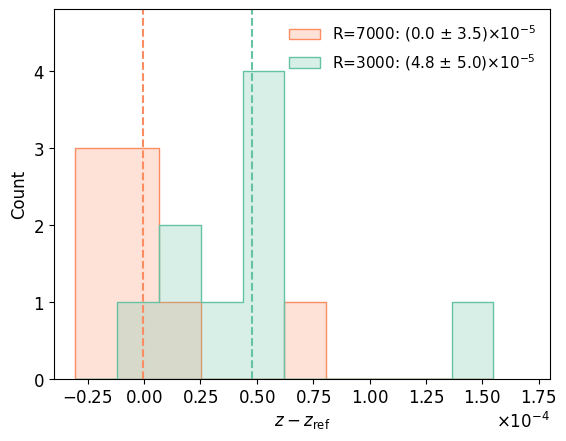}
    \caption{Redshift offset from our observations compared to the published radial velocities of radial velocity standards. Again the higher resolution performs better, but both suffer from a large positive outlier.}
    \label{fig:WiFeSRVres}
\end{figure}

\begin{figure}[bt!]
    \centering
        \includegraphics[width=0.79\textwidth]{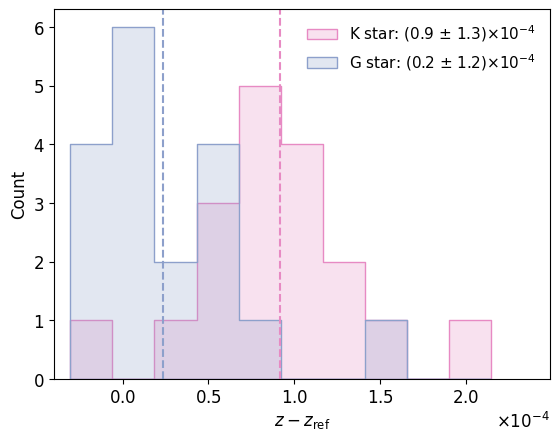}
    \caption{Redshift offset of all radial velocity standards regardless of resolution, but with redshift measured using different stellar templates in Marz. There is a disparity between the average redshift as measured by each template spectral type.}
    \label{fig:WiFeSRVtemplate}
\end{figure}

As a final test of the wavelength solution, we also observed radial velocity standard stars.
The radial velocities of these stars are known precisely \citep{Nidever2002}, so we can compare wavelength solution in a similar way to the known zero-redshift wavelength of skylines.
Figure~\ref{fig:WiFeSRVres} shows the redshift difference of each observation of a radial velocity standard (some observations are of the same star on different nights), split by resolution mode.
As expected, $R=7000$ is much more precise, with an undetectable redshift bias, whereas $R=3000$ has an offset of $5\tten{-5}$.
Both sets have $<10$ observations so it is hard to conclude if there is any meaningful correction that needs to be made to the redshifts we obtain for other targets.
Both sets are also skewed by a large outlier.
When we investigated the outlier, we found that the arc lamp frame used to calibrate the spectrum notably influenced the redshift and that the observation $\sim$0.5 hours after astronomical twilight differed in measured redshift by 7\tten{-5} from the observation $\sim$ 2.5 hours after twilight.
These two effects are seemingly unrelated, however, so this is an interesting example of how observing conditions may have unaccounted effects on redshifts.

All the radial velocity standards we chose were spectral type G or K, so we also tested how the stellar template used affected the redshift, regardless of the actual spectral type.
Thus we measured the redshift of each star with both a G and a K-star template in Marz.
Figure~\ref{fig:WiFeSRVtemplate} shows the difference in redshift offset when each template was used.
Note that we now make no distinction between resolution, and the G template histogram is exactly the combined distribution of Figure~\ref{fig:WiFeSRVres}.
Interestingly, the K-star template resulted in moderately biased redshifts, by $+7\tten{-5}$, even if the star was itself K-type.

\begin{figure*}[ht!]
    \centering
    \includegraphics[width=0.45\textwidth]{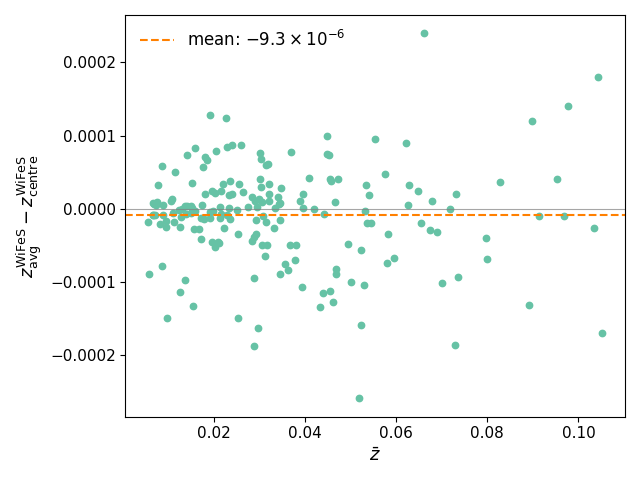}
    \includegraphics[width=0.45\textwidth]{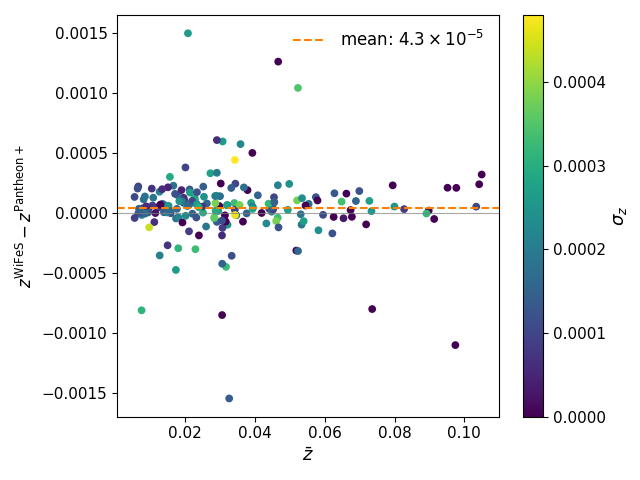}
    \caption{\textbf{Left}: Self-consistency between the average redshift over all Voronoi spaxel bins in each galaxy and just the core section. \textbf{Right}: Difference in the redshift of each galaxy as averaged over all spaxel bins from WiFeS (this work) and Pantheon+, coloured by the standard deviation in redshift over each spaxel bin within each galaxy.
    The mean offset is 4.3\tten{-5}, and insensitive to the outliers, which are interesting in themselves.
    The mean offset is similar in both panels, but the scatter in the self-consistency check in the left panel is nearly an order of magnitude smaller than the comparison with Pantheon+ redshifts in the right panel. }
    \label{fig:WiFeSzdiffs}
\end{figure*}

\begin{figure*}[t!]
    \centering
    \includegraphics[width=0.43\textwidth]{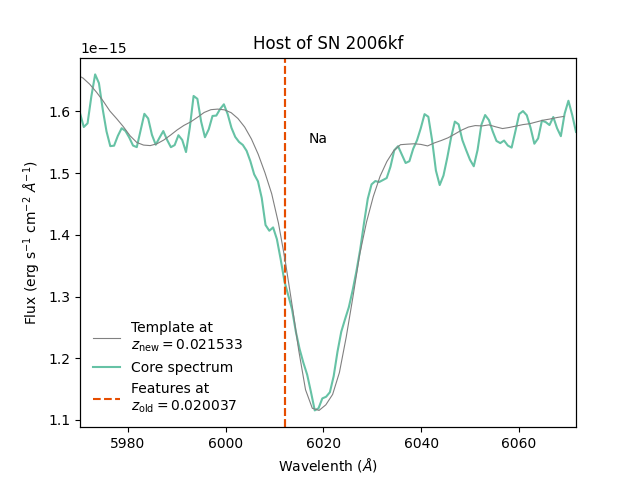}
    \includegraphics[width=0.43\textwidth]{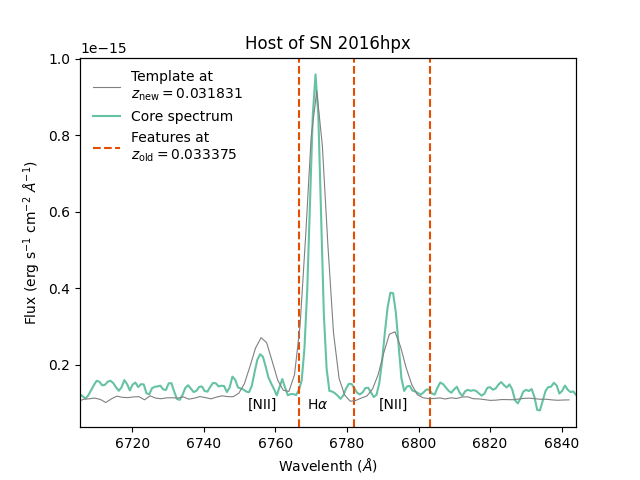}
    \caption{Comparison of WiFeS redshifts (new) to Pantheon+ redshifts (old) in the cases of the two largest discrepancies. Our redshifts are shown by our measured spectra (green), obtained at very high S/N from the core section of each galaxy, and by the grey templates, while the relevant feature locations at the Pantheon+ redshift are shown with red dashed lines.}
    \label{fig:z_comps}
\end{figure*}

Na\"{i}vely, the best solution should be to measure the redshift of a star with the closest-matching spectral type template.
However, one would also expect these two templates, in particular, to agree since the absorption features are similar for K and G.
The difference between the two templates in Marz is that the G template covers a broader wavelength range than K and includes the calcium triplet.
Assuming the broader wavelength coverage of the G-type stellar template is the main reason for the disagreement, we opted to measure the redshift of all the radial velocity standards with this template.
In addition, our redshifts from the G-star template show better agreement with the published radial velocities in every case. 

Apart from this moderate disagreement between stellar templates, none of our investigations into the accuracy of our wavelength solution revealed any need to further calibrate our redshifts.
The stellar template problem is interesting and may require further investigation, although in the case of galaxies, it remains important to measure redshifts with the template that best matches.
The redshift of a high S/N galaxy spectrum can be measured using an early, intermediate or late-type template, but the redshift may shift up to a couple of \ten{-4} depending on which main features the target and template galaxy displays.
As such, we always redshift galaxies with their matching template in Marz.

From our investigations, we are assured our redshifts are accurate to within several $\ten{-5}$. 

\section{Results}\label{sec:WiFeSresults}
We assess the success of our redshift program predominantly by estimating how accurate and precise our measurements are.
We studied the accuracy of the instrument in Section~\ref{sec:WiFeSwlcal}, so here we compare our redshifts measured from two different binning regimes (averaging over the aperture and measuring just the core) to confirm we are not biased by pointing or galaxy rotation, and we also provide a comparison to the Pantheon+ sample in Section~\ref{subsec:z_comparison}.
We study the precision of our survey in Section~\ref{subsec:z_unc} by generating many realisations of our spectra based on their measured noise and isolating individual spectral feature redshifts.
Importantly, we check how our redshifts affect cosmology in Section~\ref{subsec:cosmology}, as we specifically targeted galaxies that have the greatest potential to shift $H_0$.

\subsection{Redshift performance and comparison}\label{subsec:z_comparison}
Most galaxies (161/185) are bright and large enough in the sky to obtain at least one redshift at different spatial locations with which we can characterise the average redshift.
Others (24/185), however, were too small, dim, or otherwise had too little S/N to obtain multiple redshifts, so instead the redshift was measured using the entire spatial extent of the galaxy.
The average redshift also reflects the systemic redshift of the galaxy provided that the aperture was centred on the galaxy.
Since this is not always the case, we also measure the redshift of just the core section of the galaxy, where possible, to compare to.
This comparison is shown in Figure~\ref{fig:WiFeSzdiffs}.
The mean offset is $-9.3\tten{-6}$, and the scatter is of the order of several \ten{-5}.
The agreement on average is as good as we can expect given our investigations into the accuracy of the wavelength solution, but the scatter is also affected by whether the galaxy was centred in the aperture, the S/N, and whether the galaxy was early or late-type (roughly absorption or emission features predominantly being displayed, respectively).
The last two effects are discussed in Section~\ref{subsec:z_unc} regarding the redshift variation we might expect to see when measuring redshift from high/low S/N or narrow/wide absorption or emission features.
The largest outlier is the host of SN 2009Y, NGC 5728, which has very strongly double-peaked emission lines, even in the central region. 
In this case, using the Ca triplet to measure the redshift is much preferred, but it was not always present with high S/N.

\begin{figure*}[b!]
    \centering
    \begin{minipage}{0.49\textwidth}
        \centering
        \begin{subfigure}{0.9\textwidth}
            \includegraphics[width=0.9\textwidth]{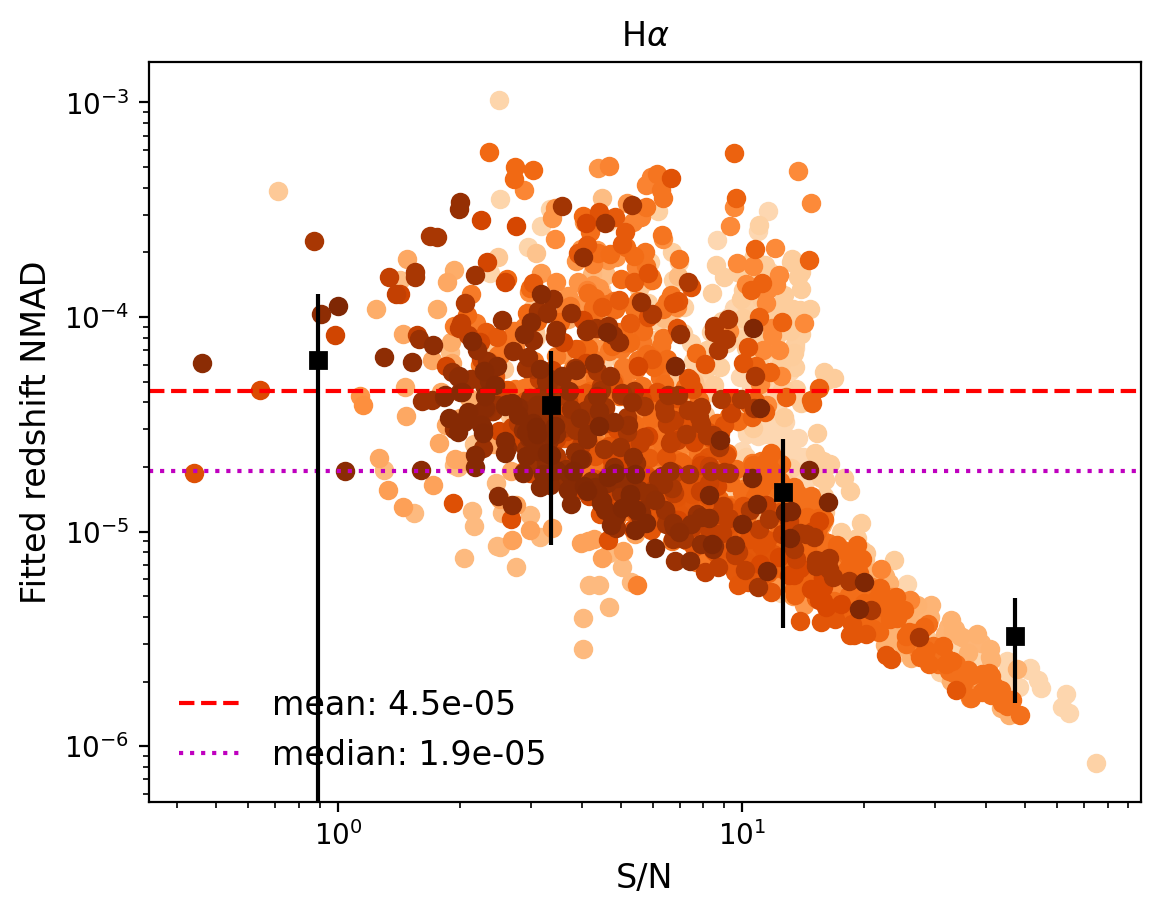}
            \includegraphics[width=0.9\textwidth]{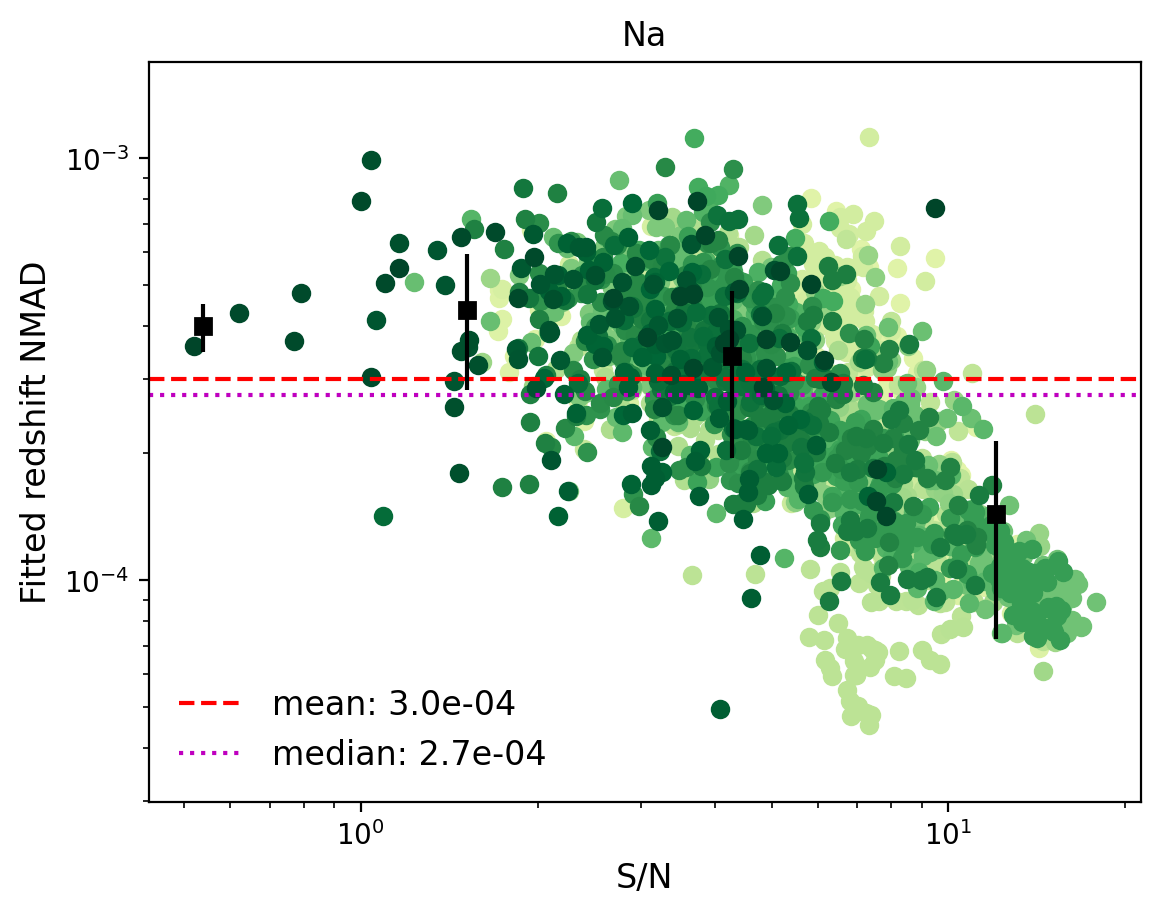}
            \caption{Examples of intra-line redshift variation, i.e.~the NMAD of the central wavelength of Gaussians fit to 500 Monte-Carlo realisations of each feature of each spectrum. The top panel shows each spaxel bin displaying H$\alpha$ emission and the bottom shows Na absorption.} \label{fig:zfits}
        \end{subfigure}
    \end{minipage}
    \begin{minipage}{0.49\textwidth}
        \centering
        \begin{subfigure}{0.92\textwidth}
            \includegraphics[width=0.9\textwidth]{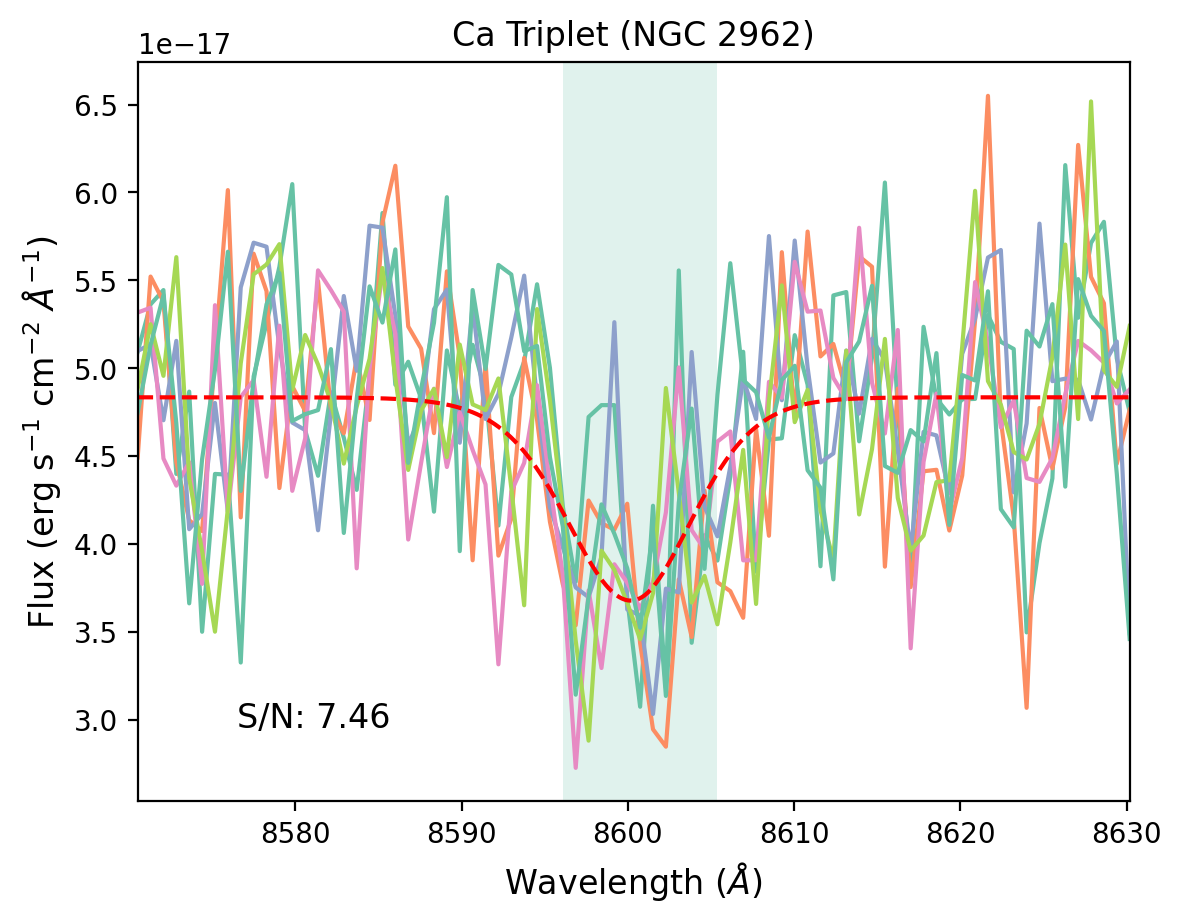}
            \caption{Example of five realisations of the Calcium triplet (central absorption line only) in NGC 2962.}\label{fig:z_MCintraline}
        \end{subfigure}
        \begin{subfigure}{0.9\textwidth}
            \includegraphics[width=0.95\textwidth]{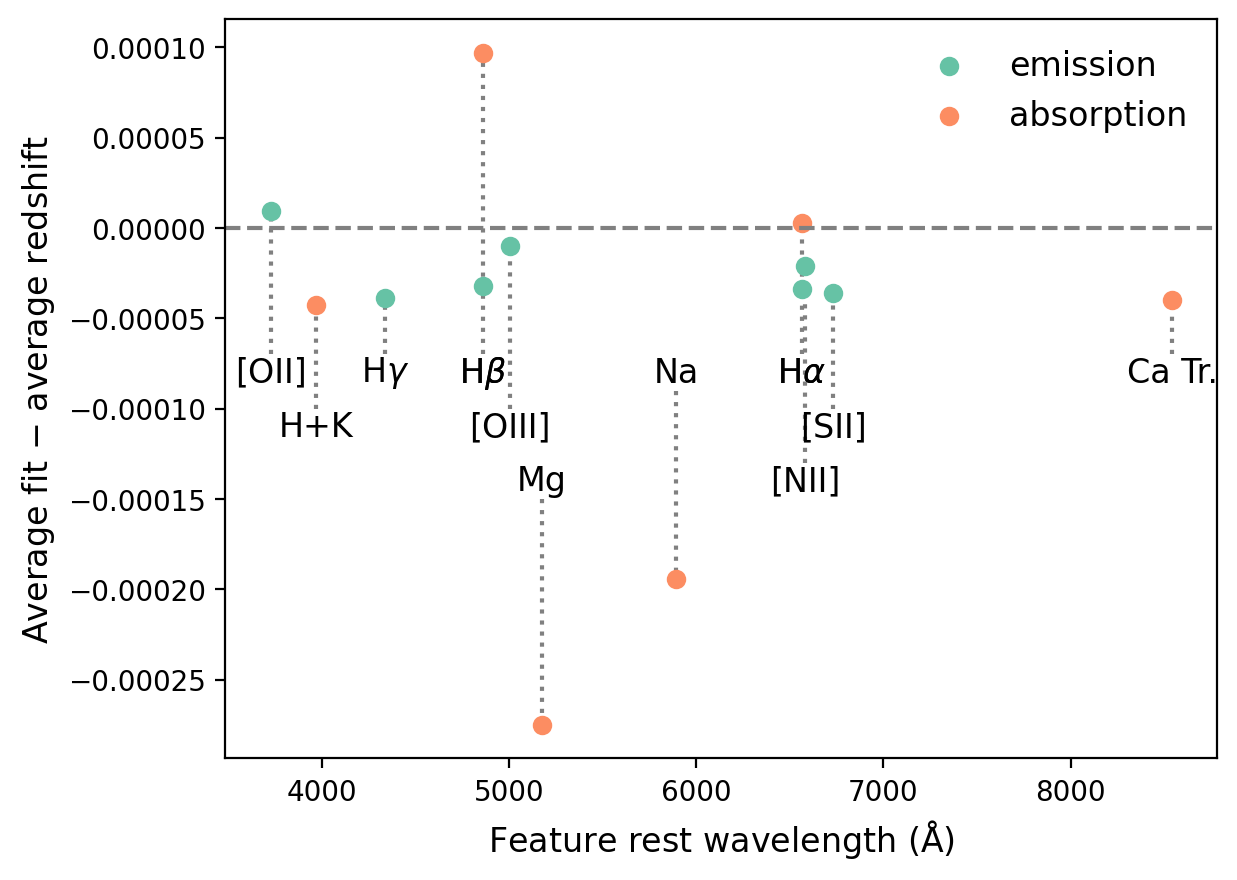}
            \caption{Variation in redshift between the mean of each feature measured individually and the corresponding spaxel bin redshift measured in the main analysis.}\label{fig:zs_interline}
        \end{subfigure}
    \end{minipage}
    \caption{\textbf{a}) shows the variation in fits to the central wavelength of the individual features of H$\alpha$ emission and Na absorption. Each spectrum was realised 500 times, varying the pixel flux by a Gaussian with a width of the measured noise, then Gaussians were fit to those features. Each point in these figures is the variation in those centres, converted to redshift, and each galaxy is a different colour. The mean and median of these are shown in the red dashed and magenta dotted lines respectively, and overplotted in black are the binned data. \textbf{b}) shows five of the 500 realisations in solid lines of a particular spaxel bin of the galaxy NGC 2962. The red dashed line shows the normalised sum of all 500 fitted Gaussians. The green window is a five \AA\ rest wavelength window about the canonical redshift used to estimate the S/N of the feature. \textbf{c}) shows the variation caused by measuring single features when compared to the redshift measured from Marz, with emission (green) and absorption (orange) differentiated (H$\alpha$ and H$\beta$ occur in both, sometimes within different regions of the same galaxy). The dotted lines simply aid in identifying each feature.}
    \label{fig:z_uncs}
\end{figure*}
We present our redshifts in Table \ref{tab:WiFeSresults} and Figure \ref{fig:WiFeSzdiffs}.
We find a mean systematic offset of 4.3\tten{-5} with normalised median absolute deviation (NMAD) of 1.2\tten{-4} between our redshifts and Pantheon+, which, as shown by \citet{Carr2022} is negligible for SN cosmology.
However, surprisingly, there are several redshift discrepancies above the level of \ten{-3}, and we show the two largest in Figure \ref{fig:z_comps}.
Neither of these examples came from optical host galaxy spectra.

For SN 2006kf, the original redshift, $z_{\text{old}}=0.020037$ came from a single-peaked 21 cm profile according to the NASA/\allowbreak IPAC Extragalactic Database,\footnote{Seen in the comment of the \href{https://ned.ipac.caltech.edu/cgi-bin/datasearch?search_type=z_id\&objid=9830\&objname=UGC\%2002829\&img_stamp=YES\&hconst=73.0\&omegam=0.27\&omegav=0.73\&corr_z=1\&of=table}{NED record}, from \citet{Springob2005}.} in contrast to our measurement of $z=0.021533$ (a difference of $1.5\tten{-3}$). 
A low S/N double horn distribution could potentially be mistaken for a single horn, and therefore could bias the redshift determination by the rotation of the galaxy.

For SN 2016hpx, the original redshift $z_{\text{old}}=0.033375$ was measured from the publicly available SN classification spectrum which showed possible host galaxy H$\alpha$ emission \citep{Foley2018, Dimitriadis2016}.
Of the two publicly available reductions of the same spectrum on the Weizmann Interactive Supernova Data Repository\footnote{\url{https://www.wiserep.org/object/9647}} (WISeREP), one does indeed show a faint peak in the wavelength region that would be consistent with a host galaxy around $z=0.033$; however, the other does not.
This could be a chance detection of a faint emission line, but it is only a single, weak feature and the spectra are low resolution and quality.
The discrepancy with our measurement of $z=0.031831$ (a difference of $1.5\tten{-3}$) would be consistent with the original redshift being mistaken as it is not a case of galactic rotation since the host, LEDA 762493, is almost face-on and the SN occurred only $3^{\prime\prime}$ from the core.

Intriguingly, the magnitude of the offset with Pantheon+ is almost exactly that of the average geocentric correction of 11.7 \kms\ ($z=3.9\tten{-5}$) we apply to our redshifts. 
Since the offset is so small, the most likely reason we see it is just due to chance.
Perhaps it could imply that the historical redshifts from Pantheon+ did not have a geocentric correction applied, but this is difficult to test, as it requires the exact observation location and time.
In any case, the individual large redshift discrepancies are potentially more interesting than this small systematic offset.

\subsection{Redshift uncertainty}\label{subsec:z_unc}
Since in most cases we have many spectra for the same galaxy and many features in those spectra, we can estimate the variation in redshift caused by noise as a measure of precision. 
We can do this in two ways: the first is to measure the variation of wavelength determination of individual features, and the second is to measure the variation between multiple features.
Both can be achieved by measuring the redshift of many realisations of each spectrum with the flux of each pixel shifted by a Gaussian whose standard deviation is the measured noise of that pixel. 
The method that utilises multiple features is more appropriate for characterising redshift uncertainty as measuring redshift from a single feature is very rarely trustworthy (unless it is particularly high S/N and/or has resolved substructure), but here we already have a tight prior on the redshift, and we are mainly interested in its variation rather than its value.
Instead of running each realisation of each spectrum through Marz, we opt to fit the features using Gaussians and convert the central wavelength to a redshift.
This method is highly scaleable (no user interface and low computation time); however, it may still be interesting to compare the robust correlation method to the simplified Gaussian fitting.
Note that we cannot simply use the width of the correlation peak given by Marz to estimate redshift uncertainty as it is at least an order of magnitude larger than our actual precision.  

Each spectrum was assigned tags for which features were present with enough S/N to be able to at least somewhat reliably fit Gaussians (both absorption and emission).
500 realisations of each spectrum were generated, and a $\pm$30 \AA\ rest wavelength window containing each feature was extracted, using the outer edges to estimate and subtract the continuum.
A Gaussian was then fit to each feature; the fit was rejected if it was more than 10 \AA\ from the known wavelength, if it was unreasonably wide or low amplitude, (accounting for the fact absorption lines are generally wider and shallower than emission), and if the amplitude was positive (negative) for emission (absorption) lines, all of which indicate a failure to capture the feature of interest.
A five \AA\ rest wavelength window around the known wavelength was also used to estimate the S/N of each feature from the original spectrum.

The features we applied this process to were: H$\alpha$, H$\beta$ (both of which can be in both emission and absorption), H$\gamma$, Na, Mg, along with the second line of each of O[II], O[III], N[II], S[II], and the CaII H+K and infrared triplet absorption features. 
The NMAD of the fitted wavelengths measures how much the redshift can shift within the bounds of the measured noise, and when plotted against S/N shows a strong trend of increasing precision with increasing S/N.
The mean of all of the fitted wavelengths when compared with the known redshift measures how much the redshift can shift according to which features are present or most prominent.
This in particular should be a more accurate reflection of the total redshift precision.
Given the type of galaxy/features and S/N, an estimate of redshift precision can be made.
Ideally this would be done on an individual redshift basis, but we save a more thorough treatment for future work.

Figure~\ref{fig:z_uncs} shows examples of the methods described above and Table~\ref{tab:z_uncs} shows the numerical results.
Figure~\ref{fig:zfits} shows the `intra-line' variation of the H$\alpha$ emission and Na absorption features against their S/N estimates, while Figure~\ref{fig:z_MCintraline} shows an example of the average Gaussian fit to the 500 realisations of the Calcium triplet of one spaxel bin of NGC 2962 (host of SN 1995D).
The S/N is just an estimate because the five \AA\ window used to estimate it is too wide for some emission lines and too narrow for some absorption lines.
Occasionally, the noise is overestimated and/or the flux is underestimated (e.g.~the cases of reduction failures), which is why we see S/N < 1 but solid feature detection.
Finally, emission lines are naturally higher S/N than absorption lines, so 1:1 comparisons between the two can be misleading.
In general though, especially at high S/N, which was the aim of this program, we see excellent precision. 
However, some features do not show a trend with S/N (such as the hydrogen absorption lines and Calcium H+K), which may be due to their presence at lower S/N in general and/or a Gaussian fit being less appropriate.

Figure~\ref{fig:zs_interline} shows the offset between the redshift via a single feature and the Marz redshift. 
These measurements are found from taking the mean of the mean offsets for each feature and spaxel bin compared to their Marz measured redshift.
The emission lines are generally in much closer agreement with the final redshift measurement; the reason Mg and Na in particular are not in agreement is due to their complex line profiles. 
Since these lines are (sometimes significantly) asymmetrical and deeper in the blue end than the red, the fitted wavelength is biased blue.
While the Gaussian fit for these lines is biased, the intra-line variation should be robust to the exact location of the centre of the Gaussian approximation.

In conclusion of this investigation, the high S/N emission line galaxy redshifts are precise to better than several \ten{-5}, the high S/N absorption galaxies and low S/N emission line galaxies are precise to better than approximately 1\tten{-4}, and the low S/N absorption galaxies are not generally present.

\begin{table}[!t]
    \centering
    \scriptsize
    \caption{Summary of spectral feature fitting precision, converted to redshift. Rest wavelengths are taken from Marz and converted from vacuum to air, except for the Calcium triplet which comes from the Vienna Atomic Line Database. For doublets and triplets, we consider only the second line.}
    \label{tab:z_uncs}
    \begin{tabular}{llrrr}
    \toprule
     Feature & Rest wavel.~(\AA) & Mean NMAD & Med.~NMAD & Diff.~from \zhel{} \\
    \midrule
    O[II] & 3727.425 & 4.2\tten{-4} & 4.2\tten{-4} &  9.5\tten{-6}\\
    CaII H+K & 3968.468 & 6.0\tten{-4} & 5.9\tten{-4} & $-$4.3\tten{-5} \\
    H$\gamma$ & 4340.469 & 2.0\tten{-4} & 1.5\tten{-4} & $-$3.9\tten{-5} \\
    H$\beta$ (em.) & 4861.325 & 1.2\tten{-4} & 7.4\tten{-5} & $-$3.2\tten{-5} \\
    H$\beta$ (abs.) & 4861.325 & 5.3\tten{-4} & 5.0\tten{-4} & 9.7\tten{-5} \\
    O[III] & 5006.843 & 1.5\tten{-4} & 1.1\tten{-4} & $-$1.0\tten{-5} \\
    Mg & 5175.3 & 5.6\tten{-4} & 5.6\tten{-4} & $-$2.8\tten{-4} \\
    Na & 5894.0 & 3.0\tten{-4} & 2.7\tten{-4} & $-$1.9\tten{-4} \\
    H$\alpha$ (em.) & 6562.80 & 4.5\tten{-5} & 1.9\tten{-5} & $-$3.4\tten{-5} \\
    H$\alpha$ (abs.) & 6562.80 & 3.3\tten{-4} & 3.0\tten{-4} & 2.9\tten{-6} \\
    N[II] & 6583.408 & 7.9\tten{-5} & 5.2\tten{-5} & $-$2.1\tten{-5} \\
    S[II] & 6730.849 & 1.5\tten{-4} & 1.2\tten{-4} & $-$3.6\tten{-5} \\
    CaII triplet & 8542.088 & 2.0\tten{-4} & 1.8\tten{-4} & $-$4.0\tten{-5} \\
    \bottomrule
    \end{tabular}
\end{table}

\subsection{Cosmological Results}\label{subsec:cosmology}
To quantify the effect of our redshift updates on cosmology, we use the entire SH0ES/Pantheon+ cosmology sample\footnote{\url{https://github.com/PantheonPlusSH0ES/DataRelease}} (photometry and Cepheid calibration) and the \textsc{Pippin}\footnote{\url{https://github.com/dessn/Pippin}} end-to-end SN cosmology analysis pipeline \citep{Hinton_and_Brout2020}.
This method allows us to calculate SH0ES/Pantheon+ distance moduli using the redshifts of this work as well as take advantage of the statistical+systematic covariance matrix $\text{C}$ for both the original and updated redshift sample. 
For each of the redshifts we remeasure, we transform to the CMB frame then recalculate peculiar velocity using \texttt{pvhub}\footnote{\url{https://github.com/KSaid-1/pvhub}} to correct to the cosmological frame (\zHD{}).
The average change in peculiar velocity was zero, but individually they varied by up to $\pm80$ \kms{} ($\pm 2.7\tten{-4}$ in redshift);\footnote{The peculiar velocity field used in \texttt{pvhub} is a discretised grid in redshift-space, and smoothed with a Gaussian kernel of width 4 \hMpc{}. The peculiar velocities only change for the galaxies that are close enough to cell walls that a small redshift change causes them to fall into a new cell.} in comparison, around 15\% of the redshift shifts are larger than these maximal peculiar velocity shifts (see the right panel of Figure~\ref{fig:WiFeSzdiffs}).

\begin{figure*}[htb!]
    \centering
    \includegraphics[width=0.42\textwidth]{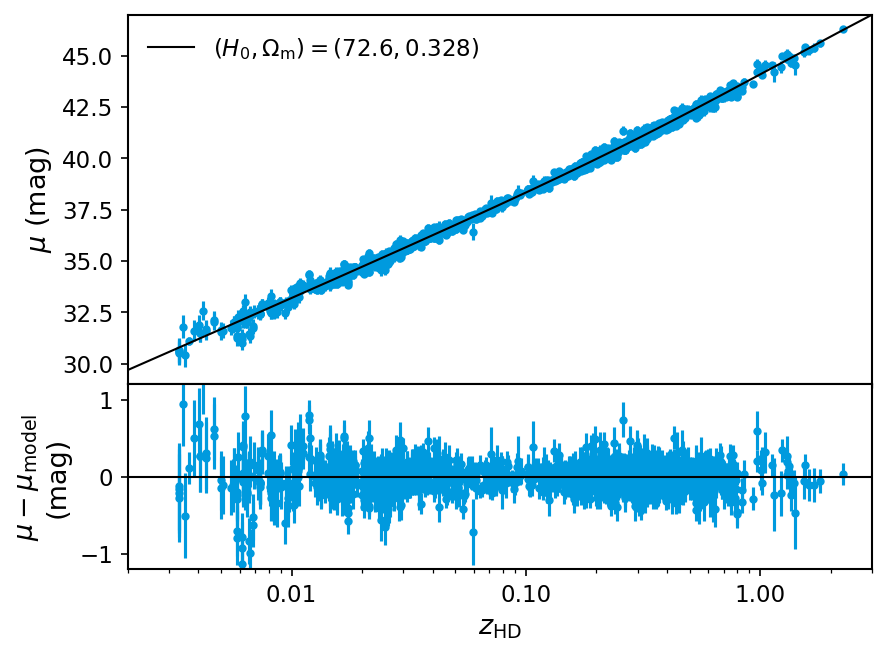}
    \includegraphics[width=0.42\textwidth]{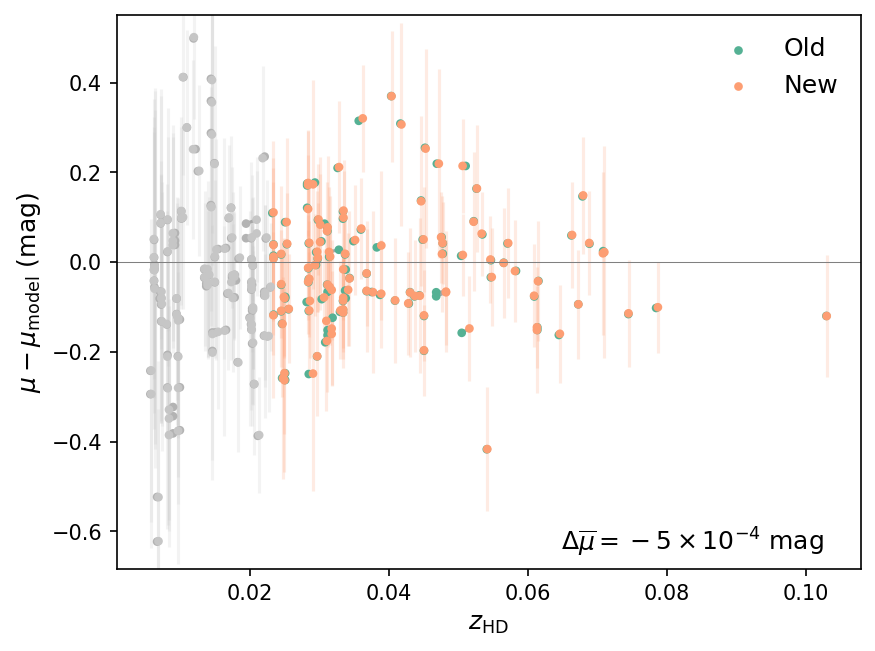}
    \caption{\textbf{Left}: full Hubble diagram showing the SNe used to constrain and the residuals of the best-fit flat $\Lambda$CDM cosmology $(H_0, \Omega_\text{m}) = (72.6$ \kmsMpc{}$, 0.328)$. \textbf{Right}: change in individual $\mu$, normalised to the best-fit cosmology of the updated redshift sample. Galaxies with $\zHD\ <0.0233$ are greyed out because they are not used to determine $H_0$. The change from original (green) to updated (orange) in magnitude space is often too small to see. The error bars, shown only for the updated sample, come from the covariance matrix statistical and systematic uncertainty added in quadrature (for display purposes only). The weighted average shift in $\mu$ of $-5\tten{-4}$ mag from the original sample corresponds to a shift in $H_0$ of $+0.1$ \kmsMpc.}
    \label{fig:H0_comp}
\end{figure*}

With our new set of cosmological redshifts \zHD{}, Pantheon+ light curves and SH0ES calibrations, we perform a simultaneous fit for $H_0$ and $\Omega_\text{m}$ in a flat $\Lambda$CDM Universe (i.e.~$\Omega_\Lambda=1-\Omega_{\text{m}}$, $w=-1$) by minimising distance modulus residuals defined by 
\begin{equation}
    \Delta \mu_i = \mu_{\text{obs},i} - \mu_{\text{model}}(z_{\text{HD},i},H_0,\Omega_{\text{m}}),
\end{equation}
where $i$ spans the set of Pantheon+ light curves.
Briefly, \textsc{Pippin} makes use of SNANA \citep{Kessler2009} to take input photometry, redshifts, etc.\ and calculate $\mu_{\text{obs},i}$ from a modified Tripp equation of the form
\begin{equation}
    \mu_{\text{obs}} = m+\alpha x_1 - \beta c - M - \delta_{\text{bias}} + \delta_{\text{host}},    
\end{equation}
where light curve peak magnitude $m$ (equivalent to $B$-band magnitude), stretch $x_1$ and colour $c$ are fit with an updated SALT2 model \citep{Guy2010, Brout2022cals}, $\alpha$ and $\beta$ are nuisance parameters, $M$ is the absolute magnitude of SNe Ia, $\delta_{\text{bias}}$ are observational bias corrections estimated from simulations, and $\delta_{\text{host}}$ accounts for the residual correlation between SN Ia brightness and host mass.
For more details, see \citet{Hinton_and_Brout2020, Brout2022cosmo}.
The theoretical distance moduli $\mu_{\text{model}}$ are calculated directly from the cosmological model as 
\begin{equation}\label{eq:dist_mod}
    \mu_{\text{model}}(z) = 5\log_{10}(D_\text{L}(z))+25,
\end{equation}
with luminosity distance $D_{\text{L}}$ in Mpc, calculated from 
\begin{equation}\label{eq:dL}
    D_{\text{L}}(z) = (1+z)\frac{c}{H_0} \int_{0}^{z} \frac{\text{d}z^\prime}{\sqrt{\Omega_{\text{m}}(1+z^\prime)^3 + 1-\Omega_{\text{m}}}}.
\end{equation}
For Cepheid calibrated galaxies, $\mu_{\text{model}}(z_i)$ is replaced with $\mu_i^{\text{Cepheid}}$.
With the vector of residuals $\Delta\boldsymbol{\mu}$, the best-fit cosmology comes from minimising the function
\begin{equation}
    \chi^2 = \Delta\boldsymbol{\mu}^\text{T}\text{C}^{-1}\Delta\boldsymbol{\mu}.
\end{equation}
Of the 185 SN host galaxies we measured, 146 galaxies and 215 light curves\footnote{Pantheon+ released 1701 light curves for 1550 unique SNe, with duplicate light curves observed by different instruments. This is accounted for in the covariance matrix, rather than combining or removing any light curves.} passed the quality cuts to be used in the fit.

This determination of $H_0$ is equivalent to fitting the intercept of the linear distance-redshift relation, as done by SH0ES \citep{Riess2022}.
The intercept, $a_\mathrm{B}$, is constrained by galaxies whose motion is dominated by expansion.
It is standard practice to use the (third-order) cosmographic expansion of recession velocity, which is almost exact in the `Hubble Flow' redshift range used for fitting $H_0$ ($0.0233<\zHD{}<0.15$):

\begin{small}\begin{align}
    \log_{10}H_0 \approx & \; \log_{10}\left[\frac{cz}{6}\left(6+3\left(1-q_0\right)z+\left(1-q_0-3q_0^2+j_0\right)z^2\right)\right] \nonumber \\ 
 & \; +\frac{1}{5}\left(-m_B + M_B + 25\right)\nonumber \\
   \equiv & \; a_B+0.2M_B+5.
\end{align}\end{small}
\noindent The expansion includes the cosmic deceleration parameter ($q_0=-0.55$) and jerk ($j_0=1.0$), whose values are chosen to match the standard $\Lambda$CDM model with $(\Omega_\text{m},\Omega_\Lambda,w) = (0.3,0.7,-1)$. 
As such, this fitting method has weak cosmological model dependence.
Since the dependence is weak, this method can still be used to constrain cosmologies whose parameters are somewhat near the input parameters.

An input cosmology also enters into our analysis in the simulations used to perform bias corrections, and this dependence is also weak \citep{Camilleri2024}.
The results of our fit are shown in Figure~\ref{fig:H0_comp}; the Hubble diagram of the sample including our redshift updates has a best-fit cosmology $(H_0, \Omega_\text{m}) = (72.6\pm1.2$ \kmsMpc{}$, 0.328^{+0.018}_{-0.017})$, and the weighted average difference in distance moduli from the sample with original redshifts is only $-5\tten{-4}$ mag, equivalent to a difference in $H_0$ of $0.1$ \kmsMpc.
The error bars we show are the statistical and systematic uncertainties from the covariance matrix, added in quadrature. 
It should be noted that even when including the 250 \kms\ uncertainty in peculiar velocities ($\sim 8\tten{-4}$ when converted to redshift), redshift uncertainties propagated through to $H_0$ are completely subdominant to these distance modulus uncertainties.
Within the SH0ES/Pantheon+ Hubble flow range, our 146 redshifts represent approximately a quarter of all SNe that have high-quality light curves.

Since we are using magnitudes calibrated with the SH0ES distance ladder, we see similar central values of $H_0$ to SH0ES \citep[$73.04\pm1.04$ \kmsMpc;][]{Riess2022}, but only the shift in $H_0$ from our redshift changes is important here.
Thus, the key takeaway from this work is that the shift in $H_0$ from the original sample purely due to our redshift updates is only $0.1$ \kmsMpc, which is negligible compared to the 1.2 \kmsMpc\ uncertainty of each fit.
We can also calculate an individual $H_0$ directly for each of the host galaxies in our sample from Equation~(\ref{eq:dL}) with $\Omega_{\text{m}}=0.3$, replacing the left-hand-side with $D_{\text{L}}(\mu_{\text{obs}})$, and taking the weighted average before and after the redshift updates gives the same result of $0.1$ \kmsMpc.
This is not unexpected from the magnitude of the changes to redshift, and a similar result was found in \citet{Carr2022} although opposite in sign.
The average redshift correction was an order of magnitude larger in \citet{Carr2022} in the same direction, so the result of this work is likely just a statistical fluctuation.
Our result reinforces the general conclusion that while redshift errors have the potential to bias $H_0$, the reality is that any realistic redshift errors are too small to affect $H_0$.

\section{Conclusion}\label{sec:WiFeSconclusion}
We have shown with our new observations that indeed there exist errors in the previous measurements of the redshifts of nearby SN Ia host galaxies.
The differences had a negligible systematic offset, which was reflected in the negligible change to $H_0$.

We thoroughly examined the instrumental accuracy with three probes in Section~\ref{sec:WiFeSwlcal}.
We tracked the wavelength solution corrections due to temperature fluctuations as measured from our frequent calibrations between science observations.
We also examined the wavelength solution accuracy across the whole aperture by checking the skyline emission spectrum on a per-night and per-run basis.
Similarly, we also compared spatially binned redshifts with accurately known radial velocity standard stars.
In no case did we find a need for corrections to our redshifts, and therefore we have shown our redshifts to be accurate to within a couple of \ten{-5}.

However, there are several extensions to our analysis that can be made to further investigate the sources of the redshift errors in the interest of mitigating them in the future.
Firstly, while we saw good agreement between averaging the redshift over all spaxels and just the core sections, it would be beneficial to investigate why we still see scatter at the level of about 3\tten{-5}.
In addition, the rest of the cases of very large redshift discrepancies (rather than just the two largest) can be examined by comparing with original spectra where possible.
Finally, for many galaxies (about half of our sample), we will be able to measure the redshift of the region around where the SN actually occurred within the galaxy.
This would be a useful crosscheck with previous redshifts, as there may be a correlation between those redshifts and the historical redshifts of Pantheon+ in the cases where a long-slit spectrograph was aligned with the SN but not the core, or a fibre-pointing inaccuracy.
It is also useful for observing the local SN environment and examining SN Ia brightness and host/environment correlations, as there is a wealth of information beyond the redshifts to explore.
As an example of how integral-field observations can be used, the high-spatial-resolution spectrograph MUSE has been used as part of the All-weather MUSE Integral-field Nearby Galaxies \citep[AMUSING;][]{Galbany2016} survey to characterise the environments of Type Ia \citep{Holoien2023} and core-collapse SNe \citep{Pessi2023}.

In this work, we study the overall accuracy and precision of our redshifts, but we have not characterised individual precisions for our redshifts, which would be a function of the spectrum S/N and number of spaxel bins for a galaxy.
Instead we give a general uncertainty class depending on the S/N and galaxy type/prominent features. 
This can be taken further to provide individual estimates, which would mostly be useful to differentiate the lower quality observations, as our high S/N data are extremely accurate and precise for an optical spectrograph without simultaneous wavelength solution measurement (from, e.g., frequency combs).
For the purposes of measuring $H_0$, our redshift data can essentially be taken as constants, at least until the peculiar velocity and distance modulus uncertainty floor is drastically improved.
This has already been done historically for convenience, but for our sample this is a valid assumption; however, it does not necessarily hold for other purposes which may be even more sensitive to redshift errors.

While we have confirmed that remeasuring accurate redshifts does not have an effect on $H_0$, we stress it is still important to use accurate and precise redshifts for cosmology, especially as we gather more and more spectroscopic data.
Importantly, at least at high-redshift, we will have to start using photometric redshifts and SN classifications, as spectroscopic follow-up becomes unfeasible from the expected volume of data from future surveys, such as the Legacy Survey of Space and Time \citep{Ivezic2019}, which will add new forms of systematic error to investigate.

\section*{Data Availability}
Data obtained as part of this work are available on Zenodo (\url{https://doi.org/10.5281/zenodo.10884817}).

\begin{acknowledgement}
The authors thank S.~Sweet for useful discussions and advice on observation optimisation, and S.~Hinton for creating the Marz redshifting program and assisting in modifying it for our purposes.
The authors thank C.~Howlett, J.~Calcino and K.~Said for assistance in carrying out observations and M.~Craigie for assistance in target selection.

TMD is the recipient of an Australian Research Council Australian Laureate Fellowship (project number FL180100168) funded by the Australian Government. 
DS is supported by DOE grants DE-SC0010007, DE-SC0021962 and the David and Lucile Packard Foundation. 
DS is supported in part by the National Aeronautics and Space Administration (NASA) under Contract No.~NNG17PX03C issued through the Roman Science Investigation Teams Programme.

This work is based on data acquired at Siding Spring Observatory with the ANU 2.3m Telescope via programs 1200040, 2200080, 3200100 and 4200059. We acknowledge the traditional custodians of the land on which the telescope stands, the Gamilaraay people, and pay our respects to elders past and present. 
This research was also supported by resources provided by the University of Chicago Research Computing Center and used services provided by the Astro Data Lab at the US National Science Foundation's National Optical-Infrared Astronomy Research Laboratory. 
NOIRLab is operated by the Association of Universities for Research in Astronomy (AURA), Inc. under a cooperative agreement with the National Science Foundation.
This work has also made use of the VALD database, operated at Uppsala University, the Institute of Astronomy RAS in Moscow, and the University of Vienna.
\end{acknowledgement}

\bibliography{references}

\clearpage
\onecolumn

\appendix
\renewcommand{\thetable}{A\arabic{table}}
\renewcommand{\theHtable}{A\arabic{table}} 
\setcounter{table}{0}

\section{Supplementary Data Tables}
\begin{ThreePartTable}
\centering
\begin{TableNotes}[flushleft]
\item [a] Exposure times are listed per observation frame. Galaxies were always observed in three frames, whereas stars were a single frame. 
\item [b] Occasionally, the same object was observed more than once on the same night. Since these data are reduced (coadded) to a single data cube, we quote the one record here but with the individual exposure times listed separately.
\item [c] SN 2005kt was in Pantheon, but not Pantheon+ as the Type Ia classification is not secure \citep{Sako2018, Carr2022}. Thus, the host and coordinates are taken from the NASA/IPAC Extragalactic Database.
\end{TableNotes}
\begin{longtable}{cllS[table-format=3.3]S[table-format=3.3,retain-explicit-plus]r}
\caption{Observing log of all science targets and radial velocity standard stars. The coordinates are those that were targeted as recorded by the instrument. The geocentric-to-heliocentric correction for each of these objects can be reconstructed with these data, along with the longitude, latitude and altitude of the telescope (149.0612$^{\circ}$, $-$31.27336$^{\circ}$, 1149.0 m respectively). \label{tab:Obslog}} \\
\toprule
& & & {Target RA} & {Target Dec} & {Exp.~time\tnote{a}} \\
{MJD} & {Target} & {SN} & {$^{\circ}$ (J2000)}  & {$^{\circ}$ (J2000)} & {s} \\
\midrule
\endfirsthead
\caption*{Observing log. (continued)}\\
\toprule
& & & {Target RA} & {Target Dec} & {Exp.~time\tnote{a}} \\
{MJD} & {Target} & {SN} & {$^{\circ}$ (J2000)}  & {$^{\circ}$ (J2000)} & {s} \\
\midrule
\endhead

\bottomrule
\endfoot

\insertTableNotes
\endlastfoot
58986.399 & ESO 125- G 006 & 2008ia & 132.635 & -61.248 & 120\\
58986.423 & WISEA J091517.24-253600.6 & 2006lu & 138.845 & -25.596 & 240\\
58986.441 & NGC 2811 & 2005am & 139.085 & -16.309 & 120\\
58986.457 & NGC 3663 & 2006ax & 170.983 & -12.318 & 120\\
58986.492 & NGC 4493 & 1994M & 187.769 & 0.593 & 120\\
58986.509 & MCG -02-34-061 & 2007ca & 187.769 & 0.593 & 150\\
58986.516 & HD130322 & \ldots & 221.886 & -0.281 & 100\\
58986.540 & UGC 08783 & AT2017cfc & 187.769 & 0.593 & 150\\
58986.563 & LEDA 766647 & 2008cf & 211.876 & -26.530 & 240\\
58986.587 & MCG -01-39-003 & 2005cf & 230.412 & -7.464 & 120\\
58986.603 & 2MASX J15453055-1309057  & ASASSN-16br & 236.375 & -13.172 & 180\\
58986.622 & 2MASX J15570808-1240252 & ASASSN-15il & 239.294 & -12.682 & 180\\
58986.640 & CGCG 082-031 & ASASSN-15nr & 261.650 & 13.900 & 150\\
58986.660 & 2MASX J17353788+0848387 & PTSS-16efw & 263.885 & 8.826 & 240\\
58986.679 & ARK 530 & ASASSN-16lg & 267.035 & 17.625 & 150\\
58986.690 & HD145897 & \ldots & 243.462 & -11.838 & 5\\
58986.709 & UGC 11128 & ASASSN-17co & 272.325 & 18.271 & 240\\
58986.723 & NGC 6805 & 2008fl & 294.241 & -37.553 & 90\\
58986.745 & ESO 284- G 032 & 2008ff & 303.505 & -44.331 & 240\\
58986.766 & ESO 107- G 004 & 2008cc & 315.917 & -67.135 & 90\\
58986.777 & WISEA J221043.94-204725.9 & 2008go & 332.661 & -20.789 & 90\\
58986.788 & UGC 11816 & 2004ey & 327.248 & 0.428 & 120\\
58987.375 & GALEXASC J090013.19-133803.5 & ASASSN-16oz & 135.034 & -13.609 & 240\\
58987.415 & KK 1524 & 2008bc & 144.690 & -63.982 & 120\\
58987.432 & ESO 570- G 020 & 2009aa & 170.929 & -22.250 & 180\\
58987.451 & ESO 440- G 001 & ATLAS17ajn & 176.081 & -28.470 & 180\\
58987.468 & NGC 4038 & 2007sr & 176.081 & -28.470 & 120\\
58987.492 & NGC 4708 & 2005bo & 192.443 & -11.103 & 150\\
58987.503 & NGC 5018 & 2002dj & 198.256 & -19.565 & 90\\
58987.514 & 2MASX J13324217-2148034 & AT2017zd & 203.158 & -21.785 & 120\\
58987.527 & NGC 5304 & 2005al & 207.492 & -30.628 & 90\\
58987.538 & NGC 5468 & 2002cr & 211.630 & -5.479 & 90\\
58987.555 & NGC 5584 & 2007af & 215.580 & -0.422 & 120\\
58987.580 & GALEXASC J151958.89+045417.3 & 2008051 & 229.987 & 4.900 & 180\\
58987.594 & UGC 10030 & 2002ck & 236.749 & -1.011 & 120\\
58987.646 & NGC 5728 & 2009Y & 220.558 & -17.303 & 150\\
58987.687 & HD145897 & \ldots & 243.462 & -11.838 & 5\\
58987.711 & \ldots & PS15bif & 305.101 & -23.745 & 240\\
58987.723 & NGC 6962 & 2002ha & 311.828 & 0.282 & 90\\
58987.743 & 2MASX J20375343+0113100 & 2006fd & 309.461 & 1.202 & 210\\
58987.762 & NGC 6956 & PSNJ2043531 & 310.960 & 12.481 & 240\\
58987.786 & WISEA J210907.40-180607.8 & ATLAS16dqf & 317.295 & -18.120 & 240\\
58987.809 & WISEA J212342.91-005034.7 & 2006oa & 320.945 & -0.863 & 210\\
58988.359 & NGC 3261 & 2008fw & 157.211 & -44.625 & 120+120\tnote{b}\\
58988.429 & GALEXASC J104848.62-201544.1 & ASASSN-16dn & 162.194 & -20.246 & 240\\
58988.501 & 2MASX J12052488-2123572 & ASASSN-16bc & 181.354 & -21.399 & 240\\
58988.524 & 2MASX J13300119-2758297 & ASASSN-16dw & 202.482 & -27.974 & 240\\
58988.543 & MRK 1337 & 2006D & 193.140 & -9.796 & 150\\
58988.560 & ESO 578- G 026 & 2007cc & 212.148 & -21.587 & 210\\
58988.581 & ESO 510- G 031 & 2007cb & 209.593 & -23.381 & 210\\
58988.634 & 2MASX J14201699-2211186 & PS16bby & 215.077 & -22.211 & 240\\
58988.661 & NGC 5728 & 2009Y & 215.077 & -22.211 & 240\\
59044.480 & 2MFGC 12594 & PS16cqa & 234.752 & 9.296 & 210\\
59044.492 & HD156026 & \ldots & 259.056 & -26.546 & 5\\
59044.617 & NGC 6928 & 2004eo & 308.230 & 9.958 & 180\\
59044.714 & UGCA 430 & ASASSN-16jf & 339.293 & -25.257 & 180\\
59044.728 & NGC 7311 & 2005kc & 338.499 & 5.553 & 180\\
59044.744 & 2MASX J22551005-0024333 & PS15bjg & 343.813 & -0.417 & 150\\
59044.773 & 2MASX J23063962-1234238 & ASASSN-15pr & 346.653 & -12.583 & 240\\
59044.788 & NGC 7721 & 2007le & 354.752 & -6.499 & 150\\
59044.800 & NGC 0191A & 2006ej & 9.760 & -9.051 & 120\\
59044.811 & NGC 0232 & 2006et & 10.650 & -23.580 & 120\\
59044.823 & UGC 00595 & 2007nq & 14.374 & -1.368 & 120\\
59045.408 & GALEXASC J134322.97-195637.5 & ATLAS17axb & 205.865 & -19.935 & 210\\
59045.431 & 2MASX J14271887-0140428 & PS16ayd & 216.847 & -1.680 & 210\\
59045.451 & SDSS J151354.30+044525.7  & ASASSN-16ct & 228.466 & 4.757 & 240\\
59045.470 & 2MFGC 12594 & PS16cqa & 234.752 & 9.296 & 210\\
59045.547 & HD156026 & \ldots & 259.053 & -26.551 & 5\\
59045.552 & HD145897 & \ldots & 243.462 & -11.837 & 5\\
59045.577 & NGC 6805 & 2008fl & 294.241 & -37.553 & 90\\
59045.598 & \ldots & PS15bif & 305.101 & -23.745 & 240\\
59045.635 & ESO 284- G 032 & 2008ff & 303.505 & -44.331 & 240\\
59045.659 & SDSS J204933.00-004543.0 & 2007ks & 312.395 & -0.784 & 240\\
59045.679 & 2MASX J21283758+0113490 & 2006eq & 322.142 & 1.225 & 180\\
59045.695 & WISEA J221440.71+050442.3 & 2007cq & 333.652 & 5.079 & 180\\
59045.710 & 2MASX J22112814-0001456 & 530086 & 332.852 & -0.020 & 150\\
59045.724 & NGC 7329 & 2006bh & 339.982 & -66.490 & 60\\
59045.745 & SDSS J224558.32-003855.9 & 2007pu & 341.480 & -0.646 & 240\\
59045.770 & WISEA J225942.70-000048.3 & 2005ku & 344.937 & -0.006 & 240\\
59045.786 & 2MASX J23154564-0120135 & ASASSN-16hz & 348.928 & -1.313 & 120\\
59045.793 & HD220957 & \ldots & 352.022 & -11.450 & 5\\
59045.803 & NGC 7780 & 2001da & 358.350 & 8.098 & 120\\
59045.816 & UGC 12859 & 2007fb & 359.219 & 5.525 & 150\\
59046.398 & 2MASX J13323577-0516218  & ASASSN-16fo & 203.132 & -5.269 & 180\\
59046.417 & IC 0986 & ASASSN-16bq & 212.874 & 1.265 & 180\\
59046.476 & HD156026 & \ldots & 259.056 & -26.546 & 5\\
59046.478 & HD156026 & \ldots & 259.056 & -26.546 & 5\\
59046.516 & UGC 10030 & 2002ck & 236.749 & -1.011 & 120\\
59046.584 & \ldots & PS15bif & 305.101 & -23.745 & 240\\
59046.596 & NGC 6962 & 2002ha & 311.828 & 0.282 & 90\\
59046.610 & ESO 107- G 004 & 2008cc & 315.917 & -67.135 & 90\\
59046.633 & 2MASX J20375343+0113100 & 2006fd & 309.461 & 1.202 & 210\\
59046.655 & WISEA J210907.40-180607.8 & ATLAS16dqf & 317.295 & -18.120 & 180\\
59046.677 & WISEA J212342.91-005034.7 & 2006oa & 320.945 & -0.863 & 240\\
59046.698 & WISEA J215558.50-010412.9 & 2006on & 329.009 & -1.076 & 180\\
59046.719 & 2MFGC 16592 & 2005lk & 329.977 & -1.202 & 210\\
59046.740 & WISEA J223041.16-004634.2 & 2005ff & 337.690 & -0.787 & 210\\
59046.763 & WISEA J224142.06-000812.7 & 2006py & 340.406 & -0.128 & 180\\
59046.782 & MCG -02-60-012 & PS15bsq & 355.477 & -8.617 & 150\\
59131.410 & NGC 6928 & 2004eo & 308.230 & 9.958 & 180\\
59131.441 & SDSS J204853.04+001129.8 & 2005fn & 312.207 & 0.192 & 240\\
59131.470 & WISEA J221225.27+005105.3 & 420100 & 333.127 & 0.852 & 150\\
59131.480 & NGC 7503 & 2001ic & 347.680 & 7.538 & 90\\
59131.485 & HD1388 & \ldots & 4.495 & -13.456 & 5\\
59131.499 & MCG -02-60-012 & PS15bsq & 355.477 & -8.617 & 150\\
59131.512 & ESO 538- G 013 & 2005iq & 359.671 & -18.722 & 120\\
59131.525 & MCG -02-01-014 & 2008hj & 0.980 & -11.176 & 150\\
59131.546 & MCG -02-02-086 & 2003ic & 10.408 & -9.280 & 150\\
59131.557 & UGC 00607 & 1999ef & 14.704 & 12.734 & 120\\
59131.568 & UGC 00595 & 2007nq & 14.374 & -1.368 & 120\\
59131.582 & WISEA J005618.02-013730.9 & 2006gt & 14.087 & -1.643 & 150\\
59131.596 & 2MASX J01135716+0022171 & 2006hx & 18.488 & 0.371 & 150\\
59131.607 & NGC 0524 & 2008Q & 21.258 & 9.574 & 120\\
59131.623 & NGC 0539 & 2008gg & 21.310 & -18.156 & 180\\
59131.641 & NGC 0692 & 2007st & 27.107 & -48.667 & 120\\
59131.656 & ESO 479- G 007 & ASASSN-16ip & 36.840 & -23.947 & 150\\
59131.671 & NGC 1015 & 2009ig & 39.594 & -1.342 & 150\\
59131.684 & NGC 1309 & 2002fk & 50.579 & -15.421 & 120\\
59131.705 & NGC 1404 & 2007on & 54.712 & -35.564 & 60\\
59131.720 & 2MFGC 03182 & 2009kk & 57.405 & -3.258 & 150\\
59131.734 & UGC 02998 & 2009ab & 64.176 & 2.755 & 150\\
59131.747 & ESO 552- G 052 & 2006hb & 75.462 & -21.129 & 150\\
59131.756 & NGC 1819 & 2005el & 77.910 & 5.191 & 90\\
59132.435 & SDSS J204933.00-004543.0 & 2007ks & 312.395 & -0.784 & 240\\
59132.458 & 2MASX J22332338-0121266 & PS16evk & 338.344 & -1.380 & 210\\
59132.467 & HD220957 & \ldots & 352.022 & -11.450 & 6\\
59132.470 & HD1338 & \ldots & 4.495 & -13.456 & 6\\
59132.484 & WISEA J232640.11-005025.9 & 2006fy & 351.662 & -0.853 & 180\\
59132.500 & GALEXASC J235326.18-153921.5 & PS15brr & 358.318 & -15.657 & 180\\
59132.522 & GALEXASC J000703.01-204149.5 & PS16fbb & 1.755 & -20.683 & 240\\
59132.542 & GALEXASC J003445.02-060936.8 & SN2016gmb & 8.683 & -6.144 & 240\\
59132.561 & 2MASX J01242239+0335168 & PS15cku & 21.093 & 3.588 & 180\\
59132.578 & UGC 00881 & 2008gl & 20.198 & 4.817 & 180\\
59132.591 & NGC 0632 & 1998es & 24.349 & 5.892 & 150\\
59132.608 & NGC 0809 & 2006ef & 31.061 & -8.717 & 180\\
59132.619 & ESO 478- G 006 & 2009le & 32.294 & -23.431 & 120\\
59132.632 & MCG -01-07-004 & ASASSN-15od & 35.816 & -4.502 & 150\\
59132.649 & UGC 02019 & 2010A & 38.180 & 0.644 & 150\\
59132.662 & IC 1844 & 1995ak & 41.455 & 3.212 & 150\\
59132.672 & NGC 1200 & 2008R & 45.912 & -11.935 & 120\\
59132.688 & MCG +00-09-074 & 2008gp & 50.760 & 1.344 & 150\\
59132.703 & UGC 02829 & 2006kf & 55.472 & 8.190 & 180\\
59132.717 & ESO 549- G 031 & 2009D & 58.626 & -19.202 & 150\\
59132.732 & NGC 1562 & ASASSN-16aj & 65.431 & -15.771 & 150\\
59132.747 & UGC 03236 & 2009ad & 75.873 & 6.670 & 180\\
59132.766 & 2MASXi J0603164-265353 & SN2016hpx & 90.830 & -26.886 & 240\\
59133.439 & 2MASX J21283758+0113490 & 2006eq & 322.142 & 1.225 & 180\\
59133.453 & WISEA J221043.94-204725.9 & 2008go & 332.661 & -20.789 & 120\\
59133.467 & WISEA J224142.06-000812.7 & 2006py & 340.406 & -0.128 & 150\\
59133.485 & WISEA J233424.11-005324.7 & 2007ra & 353.590 & -0.888 & 150\\
59133.502 & WISEA J235420.72-005501.0 & 2007om & 358.585 & -0.933 & 180\\
59133.509 & HD1388 & \ldots & 4.495 & -13.456 & 5\\
59133.511 & HD220957 & \ldots & 352.020 & -11.451 & 5\\
59133.527 & 2MASX J00343398-0112577 & 2007ht & 8.623 & -1.210 & 180\\
59133.543 & WISEA J011058.06+001634.1 & 2005kt\tnote{c} & 17.734 & 0.288 & 150\\
59133.558 & WISEA J012648.45-011417.0 & 2005hj & 21.710 & -1.253 & 180\\
59133.571 & IC 0126 & 1993ae & 22.420 & -1.976 & 150\\
59133.593 & LEDA 5069093 & 2008fr & 17.969 & 14.641 & 240\\
59133.608 & GALEXASC J013415.00-174836.1 & MASTERJ0134 & 23.561 & -17.811 & 180\\
59133.626 & NGC 0799 & 2004dt & 30.551 & -0.101 & 150\\
59133.640 & GALEXASC J021558.44+121415.2 & PS15coh & 33.993 & 12.221 & 210\\
59133.654 & 2MASX J02491020+1436036 & AT2017ns & 42.280 & 14.615 & 150\\
59133.668 & UGC 02320 NOTES01 & 2003iv & 42.560 & 12.826 & 120\\
59133.678 & CGCG 415-040 & ATLAS16dpb & 44.367 & 5.989 & 120\\
59133.691 & 2MASX J02353437-0603496  & ASASSN-15uw & 38.887 & -6.081 & 150\\
59133.706 & ESO 545- G 038 & 2005lu & 39.035 & -17.249 & 150\\
59133.720 & 2MASX J03013238-1501028 & AT2017lm & 45.375 & -14.994 & 150\\
59133.735 & MCG -01-09-006 & 2005eq & 47.228 & -7.047 & 150\\
59133.750 & 2MASX J03472342+0052316 & PS15cze & 56.829 & 0.881 & 180\\
59133.767 & CGCG 391-014 & 2007jh & 54.029 & 1.085 & 210\\
59234.480 & HD25723 & \ldots & 61.095 & -12.794 & 5\\
59234.487 & HD25723 & \ldots & 61.096 & -12.796 & 5\\
59234.511 & ESO 480-IG 021 & 2008fu & 45.627 & -24.439 & 180\\
59234.534 & \ldots & 100405 & 53.642 & -27.324 & 180\\
59234.548 & ESO 552- G 052 & 2006hb & 75.462 & -21.129 & 150\\
59234.570 & \ldots & Gaia16agf & 98.540 & -25.173 & 240\\
59234.586 & ESO 492- G 002 & 2009ag & 107.865 & -26.679 & 150\\
59234.600 & IC 0494 & 2010H & 121.573 & 1.050 & 150\\
59234.615 & NGC 2618 & 2008bi & 128.959 & 0.678 & 90\\
59234.628 & ESO 018- G 018 & 2007as & 141.910 & -80.199 & 120\\
59234.641 & NGC 2765 & 2008hv & 136.879 & 3.371 & 90\\
59234.655 & 2MASX J08325728-0351295 & MASTEROTJ08 & 128.232 & -3.879 & 120\\
59234.665 & NGC 2935 & 1996Z & 144.130 & -21.132 & 90\\
59234.729 & MRK 1337 & 2006D & 193.140 & -9.796 & 120\\
59235.553 & UGC 03738 & ASASSN-16ay & 108.061 & 7.217 & 180\\
59235.567 & UGC 03787 & 2003ch & 109.523 & 9.719 & 150\\
59235.580 & UGC 04455 & 2007bd & 127.856 & -1.219 & 150\\
59235.595 & NGC 2765 & 2008hv & 136.879 & 3.371 & 120\\
59235.611 & NGC 2986 & 1999gh & 146.132 & -21.252 & 90\\
59235.622 & NGC 2962 & 1995D & 145.260 & 5.147 & 120\\
59235.640 & WISEA J095918.72-192823.2 & 2007al & 149.842 & -19.474 & 210\\
59235.658 & CGCG 063-098 & ASASSN-15hg & 148.449 & 9.176 & 150\\
59235.675 & UGC 05586 NED02 & PS16bnz & 155.173 & -2.451 & 180\\
59235.749 & MCG -02-30-003 & ASASSN-17aj & 173.276 & -10.215 & 150+120\tnote{b}\\
59236.439 & 2MASX J02320134-2639576 & AT2016htm & 37.986 & -26.668 & 180\\
59236.455 & LEDA 170061 & ASASSN-15bc & 61.592 & -8.879 & 150\\
59236.475 & ARP 327 NED04 & 2004gc & 80.477 & 6.680 & 180\\
59236.528 & 2MFGC 04279 & PS15cwx & 78.686 & 7.025 & 180\\
59236.544 & MCG -02-16-004  & ASASSN-15ss & 93.137 & -14.231 & 180\\
59236.568 & ESO 308- G 025 & 2008bq & 100.270 & -38.054 & 150\\
59236.573 & HD52265 & \ldots & 105.075 & -5.367 & 10\\
59236.600 & ESO 561- G 018 & 2008hu & 122.248 & -18.678 & 180\\
59236.622 & 2MASX J09443215-1218233 & AT2017yk & 146.147 & -12.314 & 150\\
59236.638 & WISEA J100313.52+015343.0 & 360156 & 150.789 & 1.886 & 150\\
59236.651 & LCRS B100813.8-033156 & SN2017cjv & 152.671 & -3.775 & 120\\
59236.660 & UGC 05586 NED02 & PS16bnz & 155.173 & -2.451 & 60\\
59236.668 & NGC 3388 & 2009al & 162.818 & 8.587 & 60\\
59236.679 & NGC 3332 & 2005ki & 160.095 & 9.164 & 90\\
59236.694 & NGC 3905 & 2009ds & 177.287 & -9.751 & 120\\
59236.708 & IC 3284 & 2008ar & 186.142 & 10.833 & 150\\
59236.726 & MRK 1337 & 2006D & 193.140 & -9.796 & 150\\
59236.744 & CGCG 044-042 & ASASSN-15lg & 198.898 & 3.473 & 210\\
59236.759 & UGC 08204 & SN2017hn & 196.930 & 6.354 & 150\\
59237.440 & 2MASX J02112819-1630409 & AT2016htn & 32.853 & -16.526 & 180\\
59237.459 & \ldots & 2013go & 51.937 & -28.488 & 240+90\tnote{b}\\
59237.507 & GALEXASC J032942.01-275237.5 & 080064 & 52.409 & -27.867 & 240\\
59237.526 & 2MASX J04422451-2143312 & AT2016aj & 70.602 & -21.725 & 240\\
59237.558 & WISEA J051734.55-234659.7 & 2006is & 79.393 & -23.803 & 300\\
59237.580 & ESO 125- G 006 & 2008ia & 132.635 & -61.248 & 120\\
59237.593 & NGC 2811 & 2005am & 139.085 & -16.309 & 120\\
59237.607 & KK 1524 & 2008bc & 144.690 & -63.982 & 120\\
59237.629 & 2MASX J09583540+0044336 & PS15cms & 149.627 & 0.743 & 180\\
59237.644 & CGCG 036-091 & PS16fa & 154.799 & 4.764 & 150\\
59237.660 & NGC 3261 & 2008fw & 157.211 & -44.625 & 90\\
59237.679 & WISEA J103928.52+051101.2 & 2006al & 159.853 & 5.182 & 180\\
59237.694 & 2MASX J10480747+0010017 & PS16axi & 162.055 & 0.169 & 150\\
59237.712 & LCRS B105301.1-030602 & PS16em & 163.905 & -3.380 & 210\\
59237.726 & 2MASX J11253836+0720042 & PS17bii & 171.399 & 7.321 & 150\\
59237.739 & CGCG 071-025 & PS15aii & 191.212 & 9.750 & 150\\
59237.762 & LCRS B134713.8-024957 & PS17akj & 207.442 & -3.087 & 240\\
\bottomrule
\end{longtable}
\end{ThreePartTable}

\newpage
\begin{ThreePartTable}
\centering
\begin{TableNotes}[flushleft]
\item [a] SN 2005kt was in Pantheon, but not Pantheon+ as the Type Ia classification is not secure \citep{Sako2018, Carr2022}. Thus, the reference redshift is actually from the NASA/IPAC Extragalactic Database.
\end{TableNotes}
\begin{longtable}{llS[table-format=1.6]rS[table-format=1.6]S[table-format=1.6]S[table-format=2.1e-3]}
\caption{Redshift results. For each Type Ia SN we targeted, its Pantheon (source catalogue) ID is listed along with its host galaxy ID if it has one. Also listed is the redshift from the main analysis ($z_{\mathrm{hel}}^{\mathrm{WiFeS}}$) and the number of spaxels it was measured from ($N_z$), along with the redshift from the central region only ($z_{\mathrm{hel,centre}}^{\mathrm{WiFeS}}$). The corresponding redshift from Pantheon+ ($z_{\mathrm{hel}}^{\mathrm{Pantheon+}}$) is shown along with its difference from $z_{\mathrm{hel}}^{\mathrm{WiFeS}}$. \label{tab:WiFeSresults}} \\
\toprule
{SN} & {Host} & {$z_{\mathrm{hel}}^{\mathrm{WiFeS}}$} & {$N_z$} & {$z_{\mathrm{hel,centre}}^{\mathrm{WiFeS}}$} & {$z_{\mathrm{hel}}^{\mathrm{Pantheon+}}$} & {$z_{\mathrm{hel}}^{\mathrm{WiFeS}}-z_{\mathrm{hel}}^{\mathrm{Pantheon+}}$}  \\
\midrule
\endfirsthead
\caption*{Redshift results. (continued)}\\
\toprule
{SN} & {Host} & {$z_{\mathrm{hel}}^{\mathrm{WiFeS}}$} & {$N_z$} & {$z_{\mathrm{hel,centre}}^{\mathrm{WiFeS}}$} & {$z_{\mathrm{hel}}^{\mathrm{Pantheon+}}$} & {$z_{\mathrm{hel}}^{\mathrm{WiFeS}}-z_{\mathrm{hel}}^{\mathrm{Pantheon+}}$}  \\
\midrule
\endhead

\bottomrule
\endfoot

\insertTableNotes
\endlastfoot
1993ae & IC 0126 & 0.019776 & 15 & 0.019780 & 0.019667 & 1.1e-04\\
1994M & NGC 4493 & 0.023161 & 23 & 0.023160 & 0.023083 & 7.8e-05\\
1995ak & IC 1844 & 0.022815 & 71 & 0.022730 & 0.022699 & 1.2e-04\\
1995D & NGC 2962 & 0.006558 & 23 & 0.006550 & 0.006561 & -3.2e-06\\
1996Z & NGC 2935 & 0.007548 & 72 & 0.007540 & 0.007565 & -1.7e-05\\
1998es & NGC 0632 & 0.010633 & 23 & 0.010620 & 0.010571 & 6.2e-05\\
1999ef & UGC 00607 & 0.038941 & 16 & 0.038930 & 0.038857 & 8.4e-05\\
1999gh & NGC 2986 & 0.007743 & 37 & 0.007710 & 0.007705 & 3.8e-05\\
2001da & NGC 7780 & 0.017335 & 15 & 0.017330 & 0.017381 & -4.6e-05\\
2001ic & NGC 7503 & 0.044123 & 21 & 0.044130 & 0.044089 & 3.4e-05\\
2002ck & UGC 10030 & 0.029827 & 29 & 0.029815 & 0.029742 & 5.8e-05\\
2002cr & NGC 5468 & 0.009452 & 32 & 0.009470 & 0.009417 & 3.5e-05\\
2002dj & NGC 5018 & 0.009375 & 35 & 0.009400 & 0.00937 & 5.1e-06\\
2002fk & NGC 1309 & 0.007185 & 79 & 0.007180 & 0.007185 & -4.4e-07\\
2002ha & NGC 6962 & 0.014109 & 67 & 0.014035 & 0.01407 & 6.6e-05\\
2003ch & UGC 03787 & 0.028702 & 17 & 0.028890 & 0.02862 & 8.2e-05\\
2003ic & MCG -02-02-086 & 0.055435 & 31 & 0.055340 & 0.055359 & 7.6e-05\\
2003iv & UGC 02320 NOTES01 & 0.034504 & 22 & 0.034520 & 0.03426 & 2.4e-04\\
2004dt & NGC 0799 & 0.019505 & 19 & 0.019550 & 0.019418 & 8.7e-05\\
2004eo & NGC 6928 & 0.015791 & 111 & 0.015795 & 0.015464 & 3.0e-04\\
2004ey & UGC 11816 & 0.015832 & 13 & 0.015750 & 0.015834 & -1.7e-06\\
2004gc & ARP 327 NED04 & 0.031471 & 16 & 0.031490 & 0.03192 & -4.5e-04\\
2005al & NGC 5304 & 0.012454 & 12 & 0.012480 & 0.0124 & 5.4e-05\\
2005am & NGC 2811 & 0.007095 & 162 & 0.007105 & 0.007899 & -8.1e-04\\
2005bo & NGC 4708 & 0.013902 & 31 & 0.013910 & 0.013896 & 5.9e-06\\
2005cf & MCG -01-39-003 & 0.006651 & 54 & 0.006660 & 0.00643 & 2.2e-04\\
2005el & NGC 1819 & 0.014835 & 64 & 0.014840 & 0.01483 & 4.5e-06\\
2005eq & MCG -01-09-006 & 0.029096 & 25 & 0.029130 & 0.028952 & 1.4e-04\\
2005ff & WISEA J223041.16-004634.2 & 0.089810 & 1 & 0.089690 & 0.08979 & 2.0e-05\\
2005fn & SDSS J204853.04+001129.8 & 0.095310 & 1 & 0.095270 & 0.0951 & 2.1e-04\\
2005hj & WISEA J012648.45-011417.0 & 0.057517 & 3 & 0.057470 & 0.057385 & 1.3e-04\\
2005iq & ESO 538- G 013 & 0.034126 & 15 & 0.034110 & 0.034043 & 8.3e-05\\
2005kc & NGC 7311 & 0.015125 & 35 & 0.015090 & 0.01509 & 3.5e-05\\
2005ki & NGC 3332 & 0.019584 & 20 & 0.019560 & 0.019458 & 1.3e-04\\
2005kt\tnote{a} & WISEA J011058.06+001634.1 & 0.065360 & 2 & 0.065380 & 0.065404 & -4.4e-05\\
2005ku & WISEA J225942.70-000048.3 & 0.045328 & 22 & 0.045440 & 0.045248 & 8.0e-05\\
2005lk & 2MFGC 16592 & 0.104400 & 1 & 0.104220 & 0.104161 & 2.4e-04\\
2005lu & ESO 545- G 038 & 0.032091 & 22 & 0.032080 & 0.032189 & -9.8e-05\\
2006al & WISEA J103928.52+051101.2 & 0.067770 & 1 & 0.067760 & 0.067802 & -3.2e-05\\
2006ax & NGC 3663 & 0.016722 & 29 & 0.016750 & 0.016495 & 2.3e-04\\
2006bh & NGC 7329 & 0.010894 & 27 & 0.010900 & 0.010767 & 1.3e-04\\
2006D & MRK 1337 & 0.008491 & 169 & 0.008433 & 0.00853 & -4.6e-06\\
2006ef & NGC 0809 & 0.017959 & 14 & 0.017940 & 0.017812 & 1.5e-04\\
2006ej & NGC 0191A & 0.020439 & 18 & 0.020360 & 0.02038 & 5.9e-05\\
2006eq & 2MASX J21283758+0113490 & 0.049472 & 12 & 0.049520 & 0.049408 & 2.6e-05\\
2006et & NGC 0232 & 0.022764 & 25 & 0.022640 & 0.022639 & 1.2e-04\\
2006fd & 2MASX J20375343+0113100 & 0.079965 & 21 & 0.080035 & 0.079948 & 5.3e-05\\
2006fy & WISEA J232640.11-005025.9 & 0.082766 & 5 & 0.082730 & 0.082734 & 3.2e-05\\
2006gt & WISEA J005618.02-013730.9 & 0.044810 & 4 & 0.044710 & 0.044799 & 1.1e-05\\
2006hb & ESO 552- G 052 & 0.015141 & 43 & 0.015140 & 0.014957 & 2.1e-04\\
2006hx & 2MASX J01135716+0022171 & 0.045520 & 6 & 0.045480 & 0.045389 & 1.3e-04\\
2006is & WISEA J051734.55-234659.7 & 0.031380 & 1 & 0.031320 & 0.0314 & -2.0e-05\\
2006kf & UGC 02829 & 0.021533 & 25 & 0.021540 & 0.020037 & 1.5e-03\\
2006lu & WISEA J091517.24-253600.6 & 0.053303 & 26 & 0.053270 & 0.0534 & -9.7e-05\\
2006oa & WISEA J212342.91-005034.7 & 0.062510 & 2 & 0.062505 & 0.062573 & -3.3e-05\\
2006on & WISEA J215558.50-010412.9 & 0.071820 & 1 & 0.071820 & 0.071915 & -9.5e-05\\
2006py & WISEA J224142.06-000812.7 & 0.057855 & 2 & 0.057930 & 0.057866 & 1.0e-04\\
2007af & NGC 5584 & 0.005482 & 29 & 0.005500 & 0.005524 & -4.2e-05\\
2007al & WISEA J095918.72-192823.2 & 0.012218 & 19 & 0.012220 & 0.012175 & 4.3e-05\\
2007as & ESO 018- G 018 & 0.017098 & 43 & 0.017140 & 0.017572 & -4.7e-04\\
2007bd & UGC 04455 & 0.031035 & 25 & 0.031100 & 0.03044 & 6.0e-04\\
2007ca & MCG -02-34-061 & 0.014104 & 32 & 0.014100 & 0.014066 & 3.8e-05\\
2007cb & ESO 510- G 031 & 0.036520 & 1 & 0.036570 & 0.036592 & -7.2e-05\\
2007cc & ESO 578- G 026 & 0.029051 & 14 & 0.029040 & 0.029125 & -7.4e-05\\
2007cq & WISEA J221440.71+050442.3 & 0.025927 & 26 & 0.025840 & 0.02604 & -1.1e-04\\
2007fb & UGC 12859 & 0.017990 & 23 & 0.017920 & 0.018026 & -3.6e-05\\
2007ht & 2MASX J00343398-0112577 & 0.072853 & 3 & 0.073040 & 0.072753 & 1.0e-04\\
2007jh & CGCG 391-014 & 0.040891 & 16 & 0.040850 & 0.040744 & 1.5e-04\\
2007ks & SDSS J204933.00-004543.0 & 0.096845 & 2 & 0.096855 & 0.098 & -1.1e-03\\
2007le & NGC 7721 & 0.006756 & 49 & 0.006750 & 0.006721 & 3.5e-05\\
2007nq & UGC 00595 & 0.045223 & 32 & 0.045150 & 0.04521 & 3.1e-05\\
2007om & WISEA J235420.72-005501.0 & 0.105160 & 1 & 0.105330 & 0.10484 & 3.2e-04\\
2007on & NGC 1404 & 0.006451 & 76 & 0.006460 & 0.006248 & 2.0e-04\\
2007pu & SDSS J224558.32-003855.9 & 0.091350 & 1 & 0.091360 & 0.0914 & -5.0e-05\\
2007ra & WISEA J233424.11-005324.7 & 0.089158 & 6 & 0.089290 & 0.089163 & -4.7e-06\\
2007sr & NGC 4038 & 0.005550 & 96 & 0.005640 & 0.005417 & 1.3e-04\\
2007st & NGC 0692 & 0.021252 & 31 & 0.021250 & 0.021181 & 7.1e-05\\
2008051 & GALEXASC J151958.89+045417.3 & 0.037960 & 1 & 0.038010 & 0.03777 & 1.9e-04\\
2008ar & IC 3284 & 0.026252 & 14 & 0.026230 & 0.026173 & 7.9e-05\\
2008bc & KK 1524 & 0.014828 & 40 & 0.014825 & 0.015087 & -2.7e-04\\
2008bi & NGC 2618 & 0.013532 & 33 & 0.013630 & 0.013456 & 7.6e-05\\
2008bq & ESO 308- G 025 & 0.034448 & 22 & 0.034440 & 0.034007 & 4.4e-04\\
2008cc & ESO 107- G 004 & 0.010475 & 59 & 0.010465 & 0.010304 & 2.0e-04\\
2008cf & LEDA 766647 & 0.047290 & 1 & 0.047250 & 0.04603 & 1.3e-03\\
2008ff & ESO 284- G 032 & 0.019165 & 2 & 0.019170 & 0.019249 & -7.9e-05\\
2008fl & NGC 6805 & 0.020231 & 41 & 0.020210 & 0.01988 & 3.8e-04\\
2008fr & LEDA 5069093 & 0.039500 & 1 & 0.039480 & 0.039 & 5.0e-04\\
2008fu & ESO 480-IG 021 & 0.052121 & 29 & 0.052280 & 0.052016 & 1.0e-04\\
2008fw & NGC 3261 & 0.008522 & 79 & 0.008600 & 0.008379 & 1.4e-04\\
2008gg & NGC 0539 & 0.032091 & 18 & 0.032070 & 0.032025 & 6.6e-05\\
2008gl & UGC 00881 & 0.034226 & 15 & 0.034220 & 0.0342 & 2.6e-05\\
2008go & WISEA J221043.94-204725.9 & 0.062215 & 11 & 0.062125 & 0.062273 & -1.7e-04\\
2008gp & MCG +00-09-074 & 0.033144 & 17 & 0.033170 & 0.0335 & -3.6e-04\\
2008hj & MCG -02-01-014 & 0.037609 & 41 & 0.037680 & 0.037613 & -3.7e-06\\
2008hu & ESO 561- G 018 & 0.049940 & 11 & 0.050040 & 0.049698 & 2.4e-04\\
2008hv & NGC 2765 & 0.012743 & 107 & 0.012755 & 0.012549 & 1.8e-04\\
2008ia & ESO 125- G 006 & 0.022054 & 85 & 0.022020 & 0.021942 & 1.0e-04\\
2008Q & NGC 0524 & 0.008129 & 66 & 0.008150 & 0.008016 & 1.1e-04\\
2008R & NGC 1200 & 0.013494 & 42 & 0.013490 & 0.013296 & 2.0e-04\\
2009aa & ESO 570- G 020 & 0.027383 & 25 & 0.027380 & 0.027052 & 3.3e-04\\
2009ab & UGC 02998 & 0.011102 & 25 & 0.011120 & 0.011178 & -7.6e-05\\
2009ad & UGC 03236 & 0.028356 & 21 & 0.028400 & 0.0284 & -4.4e-05\\
2009ag & ESO 492- G 002 & 0.008731 & 67 & 0.008740 & 0.008686 & 4.5e-05\\
2009al & NGC 3388 & 0.022063 & 17 & 0.022090 & 0.022069 & -6.1e-06\\
2009ds & NGC 3905 & 0.019188 & 19 & 0.019060 & 0.01909 & 9.8e-05\\
2009D & ESO 549- G 031 & 0.025100 & 42 & 0.025250 & 0.025097 & 2.5e-06\\
2009ig & NGC 1015 & 0.008825 & 33 & 0.008820 & 0.00877 & 5.5e-05\\
2009kk & 2MFGC 03182 & 0.012505 & 17 & 0.012620 & 0.012859 & -3.5e-04\\
2009le & ESO 478- G 006 & 0.017855 & 55 & 0.017870 & 0.018149 & -2.9e-04\\
2009Y & NGC 5728 & 0.009486 & 87 & 0.009635 & 0.009743 & -1.2e-04\\
2010A & UGC 02019 & 0.020815 & 54 & 0.020860 & 0.020755 & 6.0e-05\\
2010H & IC 0494 & 0.015257 & 31 & 0.015390 & 0.015197 & 6.0e-05\\
2013go & \ldots & 0.073200 & 1 & 0.073180 & 0.074 & -8.0e-04\\
420100 & WISEA J221225.27+005105.3 & 0.097830 & 1 & 0.097690 & 0.097621 & 2.1e-04\\
530086 & 2MASX J22112814-0001456 & 0.051690 & 1 & 0.051950 & 0.052003 & -3.1e-04\\
ASASSN-15bc & LEDA 170061 & 0.036928 & 18 & 0.036850 & 0.036715 & 2.1e-04\\
ASASSN-15hg & CGCG 063-098 & 0.030056 & 19 & 0.029980 & 0.029917 & 1.4e-04\\
ASASSN-15il & 2MASX J15570808-1240252 & 0.023388 & 19 & 0.023350 & 0.023316 & 7.2e-05\\
ASASSN-15lg & CGCG 044-042 & 0.020127 & 32 & 0.020180 & 0.020151 & -2.3e-05\\
ASASSN-15nr & CGCG 082-031 & 0.023168 & 25 & 0.023150 & 0.023206 & -3.8e-05\\
ASASSN-15od & MCG -01-07-004 & 0.017637 & 55 & 0.017580 & 0.017603 & 3.4e-05\\
ASASSN-15pr & 2MASX J23063962-1234238 & 0.033301 & 17 & 0.033300 & 0.033093 & 2.1e-04\\
ASASSN-15ss & MCG -02-16-004  & 0.035624 & 27 & 0.035700 & 0.035558 & 6.6e-05\\
ASASSN-15uw & 2MASX J02353437-0603496  & 0.030388 & 12 & 0.030320 & 0.030811 & -4.2e-04\\
ASASSN-16aj & NGC 1562 & 0.030620 & 22 & 0.030630 & 0.030745 & -1.2e-04\\
ASASSN-16ay & UGC 03738 & 0.028306 & 24 & 0.028290 & 0.028343 & -3.7e-05\\
ASASSN-16bc & 2MASX J12052488-2123572 & 0.032003 & 19 & 0.031970 & 0.031939 & 6.4e-05\\
ASASSN-16bq & IC 0986 & 0.024988 & 25 & 0.024990 & 0.024935 & 5.3e-05\\
ASASSN-16br & 2MASX J15453055-1309057  & 0.028661 & 26 & 0.028700 & 0.02852 & 1.4e-04\\
ASASSN-16ct & SDSS J151354.30+044525.7  & 0.041910 & 1 & 0.041910 & 0.04191 & 0.0e+00\\
ASASSN-16dn & GALEXASC J104848.62-201544.1 & 0.012920 & 1 & 0.012920 & 0.01285 & 7.0e-05\\
ASASSN-16dw & 2MASX J13300119-2758297 & 0.034638 & 14 & 0.034610 & 0.034657 & -1.9e-05\\
ASASSN-16fo & 2MASX J13323577-0516218  & 0.029234 & 9 & 0.029250 & 0.0289 & 3.3e-04\\
ASASSN-16hz & 2MASX J23154564-0120135 & 0.025443 & 17 & 0.025410 & 0.025308 & 1.3e-04\\
ASASSN-16ip & ESO 479- G 007 & 0.017167 & 19 & 0.017180 & 0.017008 & 1.6e-04\\
ASASSN-16jf & UGCA 430 & 0.011440 & 12 & 0.011390 & 0.011441 & -1.0e-06\\
ASASSN-16lg & ARK 530 & 0.021367 & 41 & 0.021380 & 0.021171 & 2.0e-04\\
ASASSN-16oz & GALEXASC J090013.19-133803.5 & 0.030150 & 1 & 0.030110 & 0.031 & -8.5e-04\\
ASASSN-17aj & MCG -02-30-003 & 0.021444 & 20 & 0.021420 & 0.021275 & 1.7e-04\\
ASASSN-17co & UGC 11128 & 0.018357 & 35 & 0.018290 & 0.018259 & 9.8e-05\\
AT2016aj & 2MASX J04422451-2143312 & 0.067430 & 1 & 0.067460 & 0.067406 & 2.4e-05\\
AT2016htm & 2MASX J02320134-2639576 & 0.043226 & 12 & 0.043360 & 0.043313 & -8.7e-05\\
AT2016htn & 2MASX J02112819-1630409 & 0.053106 & 17 & 0.053110 & 0.053117 & -1.1e-05\\
AT2017cfc & UGC 08783 & 0.023840 & 1 & 0.023820 & 0.024027 & -1.9e-04\\
AT2017lm & 2MASX J03013238-1501028 & 0.030449 & 21 & 0.030440 & 0.030636 & -1.9e-04\\
AT2017ns & 2MASX J02491020+1436036 & 0.029373 & 11 & 0.029370 & 0.028766 & 6.1e-04\\
AT2017yk & 2MASX J09443215-1218233 & 0.046670 & 14 & 0.046760 & 0.046439 & 2.3e-04\\
AT2017zd & 2MASX J13324217-2148034 & 0.029486 & 14 & 0.029650 & 0.02947 & 1.6e-05\\
ATLAS16dpb & CGCG 415-040 & 0.022781 & 22 & 0.022790 & 0.023083 & -3.0e-04\\
ATLAS16dqf & WISEA J210907.40-180607.8 & 0.021049 & 11 & 0.021095 & 0.02117 & -1.5e-04\\
ATLAS17ajn & ESO 440- G 001 & 0.028725 & 13 & 0.028820 & 0.028706 & 1.9e-05\\
ATLAS17axb & GALEXASC J134322.97-195637.5 & 0.031580 & 1 & 0.031630 & 0.031652 & -7.2e-05\\
Gaia16agf & \ldots & 0.025285 & 13 & 0.025320 & 0.025066 & 2.2e-04\\
MASTERJ0134 & GALEXASC J013415.00-174836.1 & 0.044895 & 10 & 0.044820 & 0.044846 & 4.9e-05\\
MASTEROTJ08 & 2MASX J08325728-0351295 & 0.030521 & 18 & 0.030570 & 0.030584 & -6.3e-05\\
080064 & GALEXASC J032942.01-275237.5 & 0.066290 & 1 & 0.066050 & 0.066129 & 1.6e-04\\
100405 & \ldots & 0.103453 & 11 & 0.103480 & 0.1034 & 5.3e-05\\
PS15aii & CGCG 071-025 & 0.046510 & 23 & 0.046500 & 0.046549 & -3.9e-05\\
PS15bif & \ldots & 0.079633 & 3 & 0.079673 & 0.07937 & 2.3e-04\\
PS15bjg & 2MASX J22551005-0024333 & 0.068987 & 8 & 0.069020 & 0.068888 & 9.9e-05\\
PS15brr & GALEXASC J235326.18-153921.5 & 0.052845 & 4 & 0.052950 & 0.051804 & 1.0e-03\\
PS15bsq & MCG -02-60-012 & 0.034330 & 63 & 0.034420 & 0.034304 & -1.7e-05\\
PS15cku & 2MASX J01242239+0335168 & 0.023446 & 15 & 0.023460 & 0.023273 & 1.7e-04\\
PS15cms & 2MASX J09583540+0044336 & 0.064884 & 18 & 0.064860 & 0.06479 & 9.4e-05\\
PS15coh & GALEXASC J021558.44+121415.2 & 0.019027 & 4 & 0.019040 & 0.018837 & 1.9e-04\\
PS15cwx & 2MFGC 04279 & 0.030310 & 1 & 0.030280 & 0.030065 & 2.4e-04\\
PS15cze & 2MASX J03472342+0052316 & 0.039421 & 9 & 0.039420 & 0.039371 & 5.0e-05\\
PS16axi & 2MASX J10480747+0010017 & 0.039333 & 15 & 0.039440 & 0.039299 & 3.4e-05\\
PS16ayd & 2MASX J14271887-0140428 & 0.054097 & 7 & 0.053910 & 0.053997 & 1.0e-04\\
PS16bby & 2MASX J14201699-2211186 & 0.053550 & 24 & 0.053570 & 0.053427 & 1.2e-04\\
PS16bnz & UGC 05586 NED02 & 0.062893 & 31 & 0.062860 & 0.0627 & 1.6e-04\\
PS16cqa & 2MFGC 12594 & 0.043944 & 30 & 0.044060 & 0.043857 & 7.8e-05\\
PS16em & LCRS B105301.1-030602 & 0.069997 & 8 & 0.070100 & 0.069815 & 1.8e-04\\
PS16evk & 2MASX J22332338-0121266 & 0.054530 & 1 & 0.054550 & 0.054468 & 6.2e-05\\
PS16fa & CGCG 036-091 & 0.046043 & 40 & 0.046170 & 0.04611 & -6.7e-05\\
PS16fbb & GALEXASC J000703.01-204149.5 & 0.052183 & 12 & 0.052240 & 0.0525 & -3.2e-04\\
PS17akj & LCRS B134713.8-024957 & 0.046688 & 9 & 0.046770 & 0.046808 & -1.2e-04\\
PS17bii & 2MASX J11253836+0720042 & 0.073406 & 10 & 0.073500 & 0.073391 & 1.5e-05\\
PSNJ2043531 & NGC 6956 & 0.015513 & 43 & 0.015540 & 0.015497 & 1.6e-05\\
360156 & WISEA J100313.52+015343.0 & 0.045598 & 43 & 0.045560 & 0.045507 & 9.1e-05\\
PTSS-16efw & 2MASX J17353788+0848387 & 0.036146 & 19 & 0.036230 & 0.035573 & 5.7e-04\\
SN2016gmb & GALEXASC J003445.02-060936.8 & 0.058125 & 4 & 0.058160 & 0.058269 & -1.4e-04\\
SN2016hpx & 2MASXi J0603164-265353 & 0.031831 & 16 & 0.031770 & 0.033375 & -1.5e-03\\
SN2017cjv & LCRS B100813.8-033156 & 0.059512 & 4 & 0.059580 & 0.059528 & -1.6e-05\\
SN2017hn & UGC 08204 & 0.023897 & 23 & 0.023810 & 0.02385 & 4.7e-05\\
\bottomrule
\end{longtable}
\end{ThreePartTable}

\end{document}